\documentclass[aps,prd,twocolumn,groupedaddress, showkeys, showpacs]{revtex4}
\usepackage{graphicx,subfigure,enumerate,pstricks,latexsym,color,soul}
\usepackage[english]{babel}

\providecommand{\dif}{\mathrm{d}} 
\def\beq{\begin{equation}}
\def\eeq{\end{equation}}
\def\bea{\begin{eqnarray}}
\def\eea{\end{eqnarray}}

\def\d{\dif}

\def\SS{\Sigma}

 \def\tens{\mu}

\def\xx{x}
\def\yy{y}
\def\JJ{J}

\def\af{\zeta}

\def\Veff{V_{\rm eff}}

\def\tx{\widetilde{x}} \def\tE{\widetilde{E}}

\def\cP{\Pi}
\def\mP{P}

\def\der{|}

\newcommand{\Schw}{Schwarzschild}

\graphicspath{{pic/}}

\begin{document}

\title{  Acceleration of electric current-carrying string loop near \\
a Schwarzschild black hole immersed in an asymptotically uniform magnetic field}
\author{Arman Tursunov$^{(1,\;2)}$}
\author{Martin Kolo\v{s}$^{(1)}$}
\author{Zden\v{e}k Stuchl{\'i}k$^{(1)}$}
\author{Bobomurat Ahmedov$^{(2,\;3)}$}

\email{arman.tursunov@fpf.slu.cz}

\email{martin.kolos@fpf.slu.cz}

\email{zdenek.stuchlik@fpf.slu.cz}

\email{ahmedov@astrin.uz}

\affiliation{$^{(1)}$Institute of Physics, Faculty of Philosophy
and Science, Silesian University in Opava, Bezru{\v c}ovo
n{\'a}m.13, CZ-74601 Opava, Czech Republic}
\affiliation{$^{(2)}$Institute of Nuclear Physics, Ulughbek,
Tashkent 100214, Uzbekistan}
\affiliation{$^{(3)}$Ulugh Begh
Astronomical Institute, Astronomicheskaya 33, Tashkent 100052,
Uzbekistan}

\begin{abstract}
We study the acceleration of an electric current-carrying and
axially-symmetric string loop initially oscillating in the
vicinity of a Schwarzschild black hole embedded in an external
asymptotically uniform magnetic field. The plane of the string
loop is orthogonal to the magnetic field lines and the
acceleration of the string loop occurs due to the transmutation
effect turning in the deep gravitational field the internal energy
of the oscillating strings to the energy of their translational
motion along the axis given by the symmetry of the black hole
spacetime and the magnetic field. We restrict our attention to the
motion of string loop with energy high enough, when it can
overcome the gravitational attraction and escape to infinity. We
demonstrate that for the current-carrying string loop the
transmutation effect is enhanced by the contribution of the
interaction between the electric current of the string loop and
the external magnetic field and we give conditions that have to be
fulfilled for an efficient acceleration. The Schwarzschild black
hole combined with the strong external magnetic field can
accelerate the current-carrying string loop up to the velocities
close to the speed of light $v \sim c$. Therefore, the string loop
transmutation effect can potentially well serve as an explanation
for acceleration of highly relativistic jets observed in
microquasars and active galactic nuclei.
\end{abstract}



\maketitle


\section{Introduction}\label{intro}

Current-carrying string loop represents a simplified 1D model of
 magnetized-plasma structures \cite{Jac-Sot:2009:PHYSR4:}. The plasma
may exhibit a string-like structure arising from either dynamics
of the magnetic field lines in the plasma
\cite{Sem-Ber:1990:ASS:,Chri-Hin:1999:PhRvD:,Sem-Dya-Pun:2004:Sci:}
or thin isolated flux tubes produced in plasma
\cite{Spr:1981:AA:,Cre-Stu:2013:PhRvE:,Cre-Stu-Tes:2013:PlasmaPhys:,Cre-Stu:2014:PlasmaPhys:}.
Tension of such a string loop governs an outer barrier of the
string loop motion, while its worldsheet current introduces an
angular momentum barrier preventing the loop from collapse.

Dynamics of an axially symmetric string loop along the axis of
symmetry of Kerr black holes has been investigated in
\cite{Jac-Sot:2009:PHYSR4:,Kol-Stu:2013:PHYSR4:}, and extended
also to the case of  Kerr naked singularities
\cite{Kol-Stu:2013:PHYSR4:}. The string loop dynamics in the
spherically symmetric spacetimes has been studied for the case of
Schwarzschild---de Sitter (SdS) black holes
\citep{Kol-Stu:2010:PHYSR4:} and braneworld black holes or naked
singularities described by the Reissner-Nordstrom spacetimes
\cite{Stu-Kol:2012:JCAP:}. Such a configuration was also studied
in \citep{Lar:1994:CLAQG:,Fro-Lar:1999:CLAQG:}.

Quite recently, it has been demonstrated that small oscillations
of string loop around a stable equilibrium radii in the Kerr black
hole spacetimes can be well applied in astrophysical situations
related to the high-frequency quasiperiodic oscillations (QPOs)
observed in microquasars, i.e., binary systems containing a black
hole \cite{Stu-Kol:2014:PHYSR4:}.

On the other hand, it has been proposed in
\citep{Jac-Sot:2009:PHYSR4:} that the current-carrying string
loops could be relevant in an inverse astrophysical situation, as
a model of formation and collimation of relativistic jets in the
field of compact objects. The acceleration of jets is possible due
to the transmutation effect where the chaotic character of the
string-loop motion around a central black hole enables
transmission of the internal energy in the oscillatory mode to the
kinetic energy of the linear mode
\cite{Jac-Sot:2009:PHYSR4:,Kol-Stu:2010:PHYSR4:}. It has been
demonstrated in \cite{Stu-Kol:2012:PHYSR4:,Stu-Kol:2012:JCAP:}
that ultra-relativistic escaping velocities of string loop can be
really obtained even in spherically symmetric black hole
spacetimes. Efficiency of the transmutation effect is slightly
enhanced by the rotation of the central Kerr black hole as
demonstrated in \cite{Kol-Stu:2013:PHYSR4:}. Enhancement of the
efficiency of the transmutation process can be substantial in the
innermost parts of the Kerr naked singularity spacetimes
\cite{Kol-Stu:2013:PHYSR4:}, similarly to the acceleration process
occuring in the particle collisions \cite{Stu-Sche:2013:CLAQG:}.

The string loop accelerated by the gravitational field of
non-rotating black holes thus can potentially serve as a new model
of ultra-relativistic jets observed in microquasars and active
galactic nuclei. This is an important result since the standard
models of jet formation are based on the Blandford-Znajek effect
\cite{Bla-Zna:1977:MNRAS:} that requires rapidly rotating Kerr
black holes \cite{Pun:2001:Springer:}. The rotation of the central
black hole is an important aspect also in the alternate model of
ultra-relativistic jet formation based on the geodesic collimation
along the rotation axis of the Kerr spacetime metric
\cite{Gar-etal:2010:ASTRA:,Gar-Mar-San:2013:ApJ:}.

It is generally assumed that magnetohydrodynamics (MHD) of plasmas
in the combined gravitational and strong magnetic fields of
compact objects enables understanding of the
formation and
energetics of jets in the accreting systems orbiting the compact
objects. A magnetic field in the vicinity of the central black
hole plays an important
role in transporting of energy from
the plasma accreting into the black hole
to jets in the standard Blandford-Znajek model
\cite{Bla-Zna:1977:MNRAS:}. The theoretical and observational
data~\cite{Koo-Bic-Kun:1999:PASA:,Mil-etal:2006:NATUR:} shows that
there must exist a strong magnetic field in the close vicinity of
black holes. These magnetic fields can be generated by the
accretion of a plasma
into the black hole or if the black hole has a stellar mass it can
still contain the contribution from the field of the collapsed
precursor star \cite{Fro:2012:PHYSR4:}.

Understanding the dynamics of charged
particles in the combined
gravitational and magnetic fields is necessary
for modeling the
MHD processes. The single-particle dynamics is relevant also for
collective processes modeled in the framework of kinetic theory
\cite{Cre-Stu:2013:PhRvE:,Cre-Stu-Tes:2013:PlasmaPhys:,Cre-Stu:2014:PlasmaPhys:}.
Charged particle motion in electromagnetic fields surrounding
black hole or in the field of magnetized neutron stars has been
studied in a large variety of works for both equatorial and
general motion (see
e.g.\cite{Prasanna:1980:RDNC:,Fro-Sho:2010:PHYSR4:}). The special
class of the off-equatorial circular motion of charged particles
in combined gravitational and electromagnetic fields of compact
objects has been studied in papers
\cite{Kov-Stu-Kar:2008:CAQG:,Kov-etal:2010:CAQG:,Kop-etal:2010:ApJ:}.
Detailed analysis of the astrophysics of rotating black holes with
electromagnetic fields related to the Penrose process has been
discussed in \cite{Wagh-Dadhich:1989:PR:,Las-etal:2014:PYSR4:}.
Blandford-Znajek mechanism applied for a black hole with a
toroidal electric current was investigated in
\cite{Li-Xin:2000:PHYSR4:}. The oscillatory motion of charged
particles around equatorial and off-equatorial circular orbits
could be relevant in formation of magnetized string loops
\cite{Cre-Stu:2013:PhRvE:,Kov:2013:EPJP:}.

For the model of ultra-relativistic
jet formation based on the
transmutation effect in the string-loop dynamics, it is important
to clear up the role of the external influences, namely those
based on the cosmic repulsion and the large-scale magnetic fields.
It has been demonstrated in \cite{Stu-Kol:2012:PHYSR4:} that the
escaping string loops will be efficiently accelerated by the
cosmic repulsion behind the so called static radius
\cite{Stu:1983:BULAI:,Stu-Hle:1999:PHYSR4:} that plays an
important role also in the accretion toroidal structures
\cite{Stu-Sla-Kov:2009:CLAQG:} or motion of gravitationally bound
galaxies \cite{Stu-Sche:2011:JCAP:}. Here we shall study the role
of an asymptotically uniform external magnetic field in the
transmutational acceleration of electrically charged
current-carrying string loop in the field of non-rotating black
holes. The acceleration of the string loop is assumed along the
direction of the vector of the strength of the magnetic field.
Because of the axial symmetry of the investigated system of the
string loop and the central black hole immersed in the magnetic
field, we are able to describe the string loop motion by an
effective potential similar to those of the charged particle
motion. Using the results of our preceding paper
\cite{Tur-etal:2013:PHYSR4:}, we will show that the presence of
even weak magnetic field can sufficiently increase the possibility
and efficiency of conversion of the internal oscillatory energy of
the string loop into the kinetic energy of its linear transitional
motion.

We focus our attention to the simple
 case of the asymptotically uniform
magnetic field in order to
illustrate the large
scale role of the
magnetic field on the string loop motion. For the transmutation
effect itself, the local strength of the magnetic field is
relevant, but the subsequent motion is influenced by the large
scale structure of the magnetic field. Estimates related to the
magnitude of the magnetic field in astrophysically plausible
situations related to the magnetic field around neutron stars,
stellar black holes and supermassive black holes in the galactic
nuclei are presented in Appendix. According to these estimates,
the magnetic field can be always considered as test field, having
negligible influence of the spacetime structure.

\begin{figure}
\includegraphics[width=0.85\hsize]{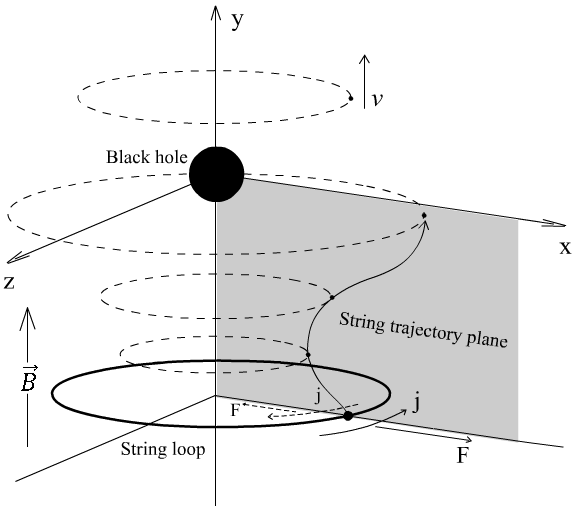}
\caption{\label{string_clas} The schematic picture of the
oscillations and the acceleration of the string loop near a black
hole embedded in an external uniform magnetic field. Due to the
axial symmetry of the system, the string trajectory can be
presented by a curve lying on the plane, chosen to be $x-y$ plane.
The direction of the Lorenz force acting on the string loop
depends on the orientation of the string loop current with respect
to the external magnetic field. The initial position of the string
loop is represented by the solid line, while its positions during
the motion are represented by the dashed lines. }
\end{figure}

The paper is organized as follows. In the Section~II, the general
relativistic description of the string loop model is presented,
the fundamental quantities that characterize the electric
current-carrying string loop through the action function and the
Lagrangian formalism are given. In Section~III, the motion of the
string loop in the combined electromagnetic and gravitational
fields of a Schwarzschild black hole immersed in an asymptotically
uniform magnetic field is studied. Due to the symmetries of the
string loop and the combined electromagnetic and gravitational
background, the string loop dynamics can be governed by a properly
defined effective potential, which is compared to the effective
potential governing motion of charged particles. At the end of the
Section~III, the physical interpretation of the string loop model
through the superconductivity phenomena of plasmas in accretion
disc is discussed. In Section~IV, the transmutation effect and the
acceleration of the string loop are studied. Dependence of the
ejection speed (relativistic Lorentz $\gamma$-factor) on the
intensity of the external uniform magnetic field is given. It is
shown that the maximal acceleration of the string loop, up to the
ultra-relativistic velocities ($v\simeq c$, $\gamma >> 1$) is
possible for the special orientation of an electric current and
the case of strong magnetic field. The concluding remarks and
discussions are presented in Section~V. In Appendix~A, the
estimation of the realistic magnetic field intensity is presented
and discussed. In Appendix~B, the dimensional analysis of the
characteristic parameters of the string loop model is presented,
along with estimates of the physical quantities that characterize
the string loop.


\section{Relativistic electric current-carrying string loop}

Generally, the string loops are assumed to be thin circular
objects that carry an current and preserve their axial
symmetry relative to the chosen axis, or the axis of the black
hole spacetime. The string loops can oscillate, changing their
radius in the loop plane, while propagating in the perpendicular
direction as shown in Figure~\ref{string_clas}. Let us consider
first a string loop moving in a spherically symmetric spacetime
with the line element
\beq
 \d s^2 = g_{tt} \d t^2  + g_{rr} \d r^2 + g_{\theta\theta} \d \theta^2 + g_{\phi\phi} \d \phi^2. \label{axsymmet}
\eeq
We will specify the components of the metric tensor
$g_{\alpha\beta}$ for the Schwarzschild black hole case in the
Section~III. The spherically symmetric Schwarzschild black hole is
assumed to be immersed in an asymptotically uniform magnetic
field; both the spacetime and the external magnetic field define
an axis of symmetry that is considered to be the axis of the
string loop.

In order to give the relativistic description for the motion of
string loop, thereby enabling the derivation of the equations of
motion one may chose the action, which will reflect the properties
of both the string loop and external fields. However, before
giving the definition of the string action, one has to introduce
the string worldsheet which is characterized by a scalar function
$\varphi$ and by coordinates $X^{\alpha}(\sigma^{a})$, where
$\alpha = 0,1,2,3$ and $a=\tau, \sigma$
\cite{Jac-Sot:2009:PHYSR4:}. The worldsheet is thus a
two-dimensional subspace which characterizes the properties of the
string loop in a given combined gravitational and electromagnetic
background. Therefore, the worldsheet of a string loop is an
analogue of the worldline of a test particle, i.e., it gives the
loci of the events of existence of the string loop in the given
background. Thus, we can introduce the worldsheet induced metric
in the following way \cite{Jac-Sot:2009:PHYSR4:}
\beq
      h_{ab}= g_{\alpha\beta}X^\alpha_{\der a}X^\beta_{\der b},
\eeq
where $X_{\der a} = \partial X /\partial a$. The current of the string loop localized on the worldsheet is described by a scalar field $\varphi$ which depends on the worldsheet coordinates $\tau$ and $\sigma$, but is independent of the choice of the spacetime coordinates. Dynamics of the string loop moving in the combined gravitational and electromagnetic field is described by the action $S$ that should contain along with the term characterizing the
freely moving string loop (see, e.g., \cite{Jac-Sot:2009:PHYSR4:}), also the term characterizing the interaction of the string loop and the external electromagnetic field. This implies the action and the Lagrangian density in the
form $\mathcal{L}$ \cite{Lar:1993:CLAQG:}:
\bea
S &=& \int \mathcal{L} \, \sqrt{-h} \, \dif \sigma \dif \tau,\\
\mathcal{L} &=& - \left[\tens/c  + \frac{k}{2} \, h^{ab} (\varphi_{|a}+A_a)(\varphi_{|b}+A_b)\right], \label{akceEM}
\eea
where we use the projection
\beq
             A_{a}=A_\gamma X^\gamma_{|a} ,
\eeq and $k$ is the constant number constrained by the world
constants. For the electrically neutral current, the constant $k$
is chosen to be equal to unity, $k=1$. The first part of
(\ref{akceEM}) represents the classical Nambu--Goto string action
with the tension $\tens$, the second part describes the scalar
field $\varphi$, rescaled according to
\cite{Jac-Sot:2009:PHYSR4:,Kol-Stu:2013:PHYSR4:}, along with the
potential $A_\alpha$ of the electromagnetic field, and their
interaction.

According to pioneering paper of Goto \cite{GOTO:1971:POTP}, the
constant $\mu/c^2$ can be interpreted as the uniform mass density
which prevents expansion of the string loop beyond some radius,
while the worldsheet current introduces an angular momentum
barrier preventing the loop from collapse. This implies that the
parameter $\mu>0$ characterizes the tension of the string loop, or
its self-force, which preserves its radius, or squeezes the string
loop. We take here the dimensions of the worldsheet coordinates as
the length dimension. Hereafter, we use the geometric units with
$c=G=1$. In these units, the constant $k$ is equal to unity as
well. The complete dimensional analysis and full conversion of the
quantities describing the string loop motion from the geometrized
units to the Gaussian or CGS units is given in the Appendix~B of
the present paper, along with estimates of the string loop
parameters in relation to realistic astrophysical conditions.

%
In the spherically symmetric spacetimes, the coordinate dependence
of the worldsheet coordinates  (\ref{axsymmet}) can be written in
the form \cite{Lar:1994:CLAQG:}
\beq
 X^\alpha(\tau,\sigma) = \left\{t(\tau),r(\tau),\theta(\tau),\sigma\right\}, \label{strcoord}
\eeq
in such a way that new coordinates satisfy the relations
\beq
 \dot{X}^\alpha = (t_{\der\tau}\,,r_{\der\tau}\,,\theta_{\der\tau}\,,0\,), \quad {X'}^\alpha = (0,0,0,1), \label{XdotXprime}
\eeq
where we use the dot to denote derivative with respect to the
string loop evolution time $\tau$, and the prime to denote
derivative with respect to the space coordinate $\sigma$ of the
worldsheet. From the relation (\ref{XdotXprime}), we can clearly
see that the string loop does not rotate in the spherically
symmetric spacetime (\ref{axsymmet}) combined with the external
uniform magnetic field.

%
The first order derivatives of the scalar field $\varphi$ with respect to the worldsheet coordinates determine the current on the string worldsheet, $\varphi_{\der a} = j_a$. The axial symmetry of the string loop model and the conformal flatness of the two-dimensional worldsheet metric $h^{ab}$ allows to write the scalar field $\varphi$ in the form  \cite{Lar:1993:CLAQG:,Jac-Sot:2009:PHYSR4:}
\beq
   \varphi =  j_{\tau} \tau + j_{\sigma}\sigma. \label{phioverj}
\eeq
In the presence of the external electromagnetic field, the equations of evolution of the scalar field will be influenced by the four-vector potential $A_\mu$; we then define the total current, labelled by the tilde $\tilde{j}_a$, in the following form
\beq
 \tilde{j}_{\tau} = j_\tau + A_\alpha X^\alpha_{\der \tau}, \quad
 \tilde{j}_{\sigma} = j_\sigma + A_\alpha X^\alpha_{\der \sigma}. \label{constJOn}
\eeq
The variation of the action (\ref{akceEM}) with respect to the scalar field $\varphi$ can now be written in the form
\beq
 \left( \sqrt{-h} \, h^{ab} \, \tilde{j}_a \right)_{\der b} = 0.  \label{PHIevolution}
\eeq
Equations of motion (\ref{PHIevolution}) for the scalar field $\varphi$, and the string loop axisymmetry imply existence of conserved quantities $\tilde{j}_{\tau}$ and ${j}_{\sigma}$. The conserved quantities $\tilde{j}_{\tau}$ and ${j}_{\sigma}$ correspond to the parameters $\Omega$ and $n$ introduced in \cite{Lar:1993:CLAQG:}, up to the  constant $k$. Quantities ${j}_{\tau}$ and $\tilde{j}_{\sigma}$ generally could not be conserved during string loop motion.

%
Varying the action (\ref{akceEM}) with respect to the four-potential $A_\alpha$ \cite{Lar:1993:CLAQG:}, we obtain the electromagnetic current density
\beq
 I^{\mu} = \frac{\delta \mathcal{L}}{\delta A_{\mu}} = k \tilde{j}_\tau \dot{X}^\mu - k \tilde{j}_\sigma {X'}^\mu,
\eeq
and we can identify the string loop electric charge density $q$,
and the electric current density $j$ due to the relations \beq
 q = k \tilde{j}_\tau, \quad j =  k \tilde{j}_\sigma. \label{QIdef}
\eeq
Up to the constant $k$, we can consider the parameters $\tilde{j}_\tau$ and $\tilde{j}_\sigma$ to be related to the  electric charge and the current densities. We can deduce from (\ref{PHIevolution}) that the electric charge density $q$ is conserved during the string loop evolution, but the current density $j$ is changing as it is influenced by the external electromagnetic field.

%
It is important to specify the worldsheet stress-energy tensor
which can be found varying the action (\ref{akceEM}) with
respect to the induced metric $h_{ab}$
 \cite{Jac-Sot:2009:PHYSR4:}
\bea
&& \SS^{\tau\tau} = \frac{k}{2} \, \frac{\tilde{j}_{\tau}^2+\tilde{j}_{\sigma}^2}{g_{\phi\phi}} + \mu, \quad
     \SS^{\sigma\tau} = - \frac{ k \tilde{j}_{\tau} \tilde{j}_{\sigma}}{g_{\phi\phi}}, \nonumber\\
&& \SS^{\sigma\sigma} = \frac{k}{2} \, \frac{\tilde{j}_{\tau}^2+\tilde{j}_{\sigma}^2}{g_{\phi\phi}} - \mu.
\eea
%

%
The string loop canonical momentum density is defined by the relation
\beq
 \cP_\mu \equiv \frac{\partial \mathcal{L}}{\partial \dot{X}^\mu} = \SS^{\tau a} g_{\mu\alpha} X^\alpha_{|a} + k \tilde{j}_\tau A_\mu . \label{canMomenta}
\eeq
From the Lagrangian density (\ref{akceEM}) we can obtain Hamiltonian of the string loop dynamics in the combined gravitational and electromagnetic field in the form \cite{Lar:1993:CLAQG:,Car-Ste:2004:PHYSR4:}
\bea
 H &=& \frac{1}{2} g^{\alpha\beta} (\cP_\alpha - q A_\alpha)(\cP_\beta - q A_\beta) \nonumber \\
   && + \frac{1}{2} g_{\phi\phi} \left[(\SS^{\tau\tau})^2 - (\SS^{\tau\sigma})^2 \right] . \label{HamEM}
\eea
The string loop dynamics is determined by the Hamilton equations
\beq
 \mP^\mu \equiv \frac{\d X^\mu}{\d \af} = \frac{\partial H}{\partial \cP_\mu}, \quad
 \frac{\d \cP_\mu}{\d \af} = - \frac{\partial H}{\partial X^\mu}. \label{Ham_eqEM}
\eeq
The canonical momentum $\cP^\mu$ (\ref{canMomenta}) is related to the mechanical momentum $\mP^\mu$ by the relation
\beq
  \cP^\mu = \mP^\mu + q A^\mu,
\eeq
where we use the string loop charge density definition (\ref{QIdef}). In the Hamiltonian (\ref{HamEM}), we use an affine parameter $\af$, related to the worldsheet coordinate $\tau$ by the transformation $\d \tau = {\SS^{\tau\tau}} \d \af$.



\section{Dynamics of string loop in uniform magnetic field around Schwarzschild black hole}

In this section we apply the previous solutions for the case of
the Schwarzschild black hole spacetime immersed in axially
symmetric magnetic field that is uniform at the spatial infinity.
The plane of the string loop is perpendicular to the lines of
strength of the magnetic field. The string loop moves along the
axis which is chosen to be $y$-axis as shown in
Figure~\ref{string_clas}. The oscillations of the string loop are
restricted to the $x$-$z$ plane (considered in the $x$-axis due to
the axisymmetry of the string loop), while its trajectory, because
of the symmetry, is considered in the $x$-$y$ plane.
The Schwarzschild black hole spacetime characterized by the mass parameter $M$ takes in the standard spherical coordinates the form
\begin{equation} \label{metric}
 ds^2=-f(r) dt^2+f^{-1}(r) dr^2+r^2 \left(d\theta^2+\sin^2\theta d\phi^2\right)\ ,
\end{equation}
where the metric "lapse" function $f(r)$ is defined by
\begin{equation}
  f(r) = 1 - \frac{2 M}{r}.
\end{equation}
{Due to the symmetries discussed above for the description of the string loop motion, it is convenient for the proper description of the string loop motion to work with the Cartesian coordinates defined as \cite{Jac-Sot:2009:PHYSR4:,Kol-Stu:2010:PHYSR4:}
\begin{equation}
 x = r \sin\theta \ , \quad y = r \cos\theta\ . \label{ccord}
\end{equation}

In our study we assume the static, axisymmetric and asymptotically
uniform magnetic field. Since the Schwarzschild spacetime is flat
at spatial infinity, the timelike $\xi_{(t)}$ and spacelike
$\xi_{(\phi)}$ Killing vectors satisfy the equations
$\Box\xi^{\alpha}=0$, which exactly correspond to the Maxwell
equations
\beq \Box A^{\alpha}=0, \eeq
for the four-vector potential of the electromagnetic field. The solution of the Maxwell equations can be then written in the Lorentz gauge in the form~\cite{Wald:1974:PHYSR4:}
\begin{equation}
A^{\alpha}=C_{1}\xi^{\alpha}_{(t)}+C_{2}\xi^{\alpha}_{(\phi)}\ .
\end{equation}
The first integration constant has to be $C_1=0$, because of the asymptotic properties of the Schwarzschild spacetime (\ref{metric}), while the second integration constant takes the form $C_2 = B/2$, where $B$ is the
strength of the homogeneous magnetic field at the spatial infinity. The commuting Killing vector $\xi_{(\phi)} = \partial / \partial \phi$ generates rotations around the symmetry axis. Consequently, the only nonzero covariant component of the potential of the electromagnetic field takes the form~\cite{Wald:1974:PHYSR4:}
\begin{equation}
A_{\phi} = \frac{B}{2} r^2 \sin^2 \theta = \frac{B}{2} x^2\ .
\label{aasbx}
\end{equation}
The symmetries of the considered background gravitational and
magnetic fields, corresponding to the $t$ and $\phi$ components of
the Killing vector, imply the existence of two constants of the
string loop motion
\cite{Tur-etal:2013:PHYSR4:,Jac-Sot:2009:PHYSR4:}
%
\bea
E &=& -\xi^{\mu}_{(t)} \cP_{\mu} = - \cP_{t}, \label{intE} \\
L &=& \xi^{\mu}_{(\varphi)} \cP_{\mu}
= - \tilde{j}_{\tau} \tilde{j}_{\sigma} + q A_{\varphi}
= - \tilde{j}_{\tau} j_{\sigma}. \label{intL}
\eea
where we have already set $k=1$. The angular momentum $L$ is given by two another constants of motions $\tilde{j}_{\tau}$ and $j_{\sigma}$.

\begin{figure}
\includegraphics[width=0.7\hsize]{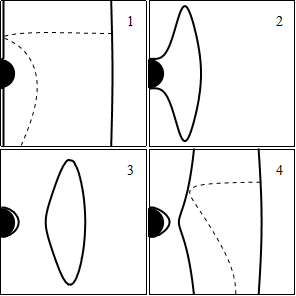}
\caption{ Four possible types of the boundaries of the string loop
motion in the Schwarzschild black hole spacetime and examples of
string loop trajectories escaping to the infinity along $y$-axis.
\label{cls} }
\end{figure}

In the spherically symmetric spacetime (\ref{metric}), the Hamiltonian (\ref{Ham_eqEM}) governing the string loop dynamics can be expressed in the form \cite{Tur-etal:2013:PHYSR4:}
\beq
 H = \frac{1}{2} f(r) \mP_r^2 + \frac{1}{2r^2} \mP_\theta^2 - \frac{E^2}{2f(r)}
     + \frac{\Veff}{2 f(r)}\ , \label{HamHam}
\eeq
with an effective potential for the string loop motion in the combined gravitational and magnetic fields
\beq
\Veff = f(r) \left\{ \frac{B^2 x^3}{8} + \left(\frac{\Omega J B}{\sqrt{2}} + \mu \right) x + \frac{J^2}{x} \right\}^2 . \label{Sveff}
\eeq
In accord with \cite{Jac-Sot:2009:PHYSR4:}, we have introduced new parameters that are conserved for the string loop dynamics in the Schwarzschild spacetime combined with the uniform magnetic field
\beq
J^2 \equiv \frac{j_{\sigma}^2 +j_{\tau}^2}{2}, \quad \omega \equiv -\frac{j_\sigma}{j_\tau}, \quad
\Omega \equiv \frac{ - \omega}{\sqrt{1+\omega^2}} , \label{Omegapar}
\eeq
where the parameter $J$ is always positive, $J>0$, the dimensionless parameter $\omega$ runs in the interval $-\infty<\omega<\infty$, and the dimensionless parameter $\Omega$ varies in the range $-1<\Omega<1$.

For the uniform magnetic field ($A_t = 0$), we can introduce relations between the string loop charge $q$ and the current $j$ densities (\ref{QIdef}), and the string loop parameters $J$,$\Omega$, in the form
\beq
q = j_\tau, \quad j = j_\sigma + A_\phi = j_\sigma  + \frac{B}{2} \, x^2, \quad  j_\sigma = \sqrt{2} J \Omega. \label{eletricQJ}
\eeq
It is worth to recall that $j_\tau$ and $j_\sigma$ are conserved quantities in the uniform magnetic fields.

Note that in absence of the external magnetic field, $B=0$, there
exists a symmetry which allows for the interchange $\omega
\leftrightarrow 1/\omega$; then the interval $-1<\omega<1$ covers
all possible cases of the string loop motion
\cite{Jac-Sot:2009:PHYSR4:,Kol-Stu:2010:PHYSR4:}. In the case of
non-vanishing magnetic field, such a symmetry does not exist, and
$\omega$ has to range generally in the interval
$(-\infty,\infty)$. The sign of the parameter $\Omega$ depends on
the choice of the direction of the electric current with respect
to the direction of the uniform magnetic field. The case
$\Omega=0$ corresponds to the zero current. The case with $\Omega
> 0$, in principle is unstable, since even a small deviation of
the string loop from the symmetry axis leads to the appearance of
a torque proportional to the product of the current and the
magnetic field and potentially the string loop can be overturned
to the stable configuration with $\Omega < 0$
\cite{Tur-etal:2013:PHYSR4:}.

The condition $H=0$ determines the regions allowed for the string loop motion, and it implies the relation for the effective potential (or energy boundary function) governing the motion through the string loop and the background (spacetime and magnetic field) parameters
\beq
 E^2 = \Veff(x,y;J,B,\Omega). \label{StringEnergy}
\eeq
Detailed analysis of the boundaries of the string loop motion
has been done in \cite{Tur-etal:2013:PHYSR4:}, we only recall main
results here. The string loop motion boundaries and related types
of the motion can be distinguished into four classes according to
the possibility of the string loop to escape to infinity or
collapse to the central compact object. The first class of the
boundaries correspond to the absence of inner and outer
boundaries, i.e. the string loop can be captured by the black hole
or escapes to infinity. The second class corresponds to the
situation with an outer boundary - the string loop must be
captured by the black hole. The third class corresponds to the
situation when both inner and outer boundaries exist - the string
loop is trapped in some region forming a potential ``lake'' around
the black hole. The fourth class corresponds to an inner boundary
- the string loop cannot fall into the black hole but it must
escape to infinity, see Fig. \ref{cls} for the details. For the
purposes of the present paper, the first and fourth classes of
boundaries, when the string loop can escape to infinity, are
relevant.


\subsection{Analogy with test particle motion}

The influence of electromagnetic field on the string loop dynamics can be cleared up, if we compare the effective potential of the string loop dynamics with those of the charged test particle motion on circular orbits around a Schwarzschild black hole immersed in the same magnetic field $B$. Hamiltonian of the motion of a charged test particle with mass $m$ and charge $q$ is given by the relation \cite{Mis-Tho-Whe:1973:Gra:}
\beq
  H_{\rm p} =  \frac{1}{2} g^{\alpha\beta} (\cP_\alpha - q A_\alpha)(\cP_\beta - q A_\beta) + \frac{1}{2} \, m^2, \label{particleHAM}
\eeq
where the mechanical and canonical momenta are again related by
\beq
\mP^\mu = \cP^\mu - q A^\mu.
\eeq
We suppose the particle moving on a circular orbit at radius $x$, with constant angular velocity $\omega = \mP^\phi/m$.

In the homogeneous magnetic field $B$, the Hamiltonian of the test particle motion (\ref{particleHAM}) can be cast into the (\ref{HamHam}) form, but with modified effective potential containing a constant axial angular momentum $L$
\beq
V_{\rm eff}^{\rm (p)} = f(r) \left\{ \frac{L}{x} - \frac{q B}{2} x \right\}^2 + m^2. \label{Pveff}
\eeq
The effective potentials $V_{\rm eff}$ for both string loop (\ref{Sveff}) and test particle (\ref{Pveff}) motion consist of the part given by the geometry, $f(r)$, and the part in braces depending only on the $x$ coordinate. We can compare terms in braces with the same $x$ dependence for both string and particle cases revealing their interpretation.

The first and the second term in braces in the effective potential of the particle motion (\ref{Pveff}) represent the angular momentum contribution, $\sim x^{-1}$, and the Lorentz force contribution, $\sim x$, respectively.

The first term in braces in the effective potential of the string loop motion (\ref{Sveff}) represents  the pure contribution of the external magnetic field to the "effective" energy - density of the magnetic field energy is proportional to $B^2$, and the space volume to $x^3$.
The second term in braces in (\ref{Sveff}), showing the same
dependence, $\sim x$, as the Lorentz force, consists from two
parts - the pure tension $\mu$, and the part representing
interaction between the electric current carried by the string
loop and the external magnetic field. This term can be associated
to interaction of two magnetic fields, where one of them is the
"global", external, magnetic field, $\sim B$, and the other is the
"local", self-generated, magnetic field related to the string
loop, $\sim J\Omega/x$, in accord with (\ref{Omegapar}). The sign
of this term can be either positive or negative, depending on the
sign of $\Omega$, i.e., the direction of the electric current on
the string loop. We consider $\Omega$ to be negative, if the
direction of the vector of the self-generated magnetic field
coincides with the direction of the vector of the external
magnetic field, and positive, if these vectors are directed
oppositely. Since the direction of the external magnetic field is
given as an initial condition, the direction of the Lorenz force
acting on the string is determined by the direction of the
current.
The last term in braces of the effective potential (\ref{Sveff})
corresponds to the angular momentum generated by the current of
the string loop.



\subsection{Analogy with the superconducting string} \label{analogy}

The string loop model with the current generated by the scalar
field $\varphi$ demonstrates an interesting analogy with the
superconductivity. In the pioneering works
\cite{Wit:1985:NuclPhysB:,Vil-She:1994:CSTD:}, it has been shown
for the cosmic strings that in the case when the electromagnetic
gauge invariance is broken, the string can be considered as a
superconductor carrying large currents and charges, up to the
order of the string mass scale. Under such circumstances, the
carriers of the electric charge can be either bosons or fermions,
depending on the energetic favor for the charged particle
\cite{Kim:1999:JKPS}. In a series of papers
\cite{Mart-Shel:1997:PRB,Car:1990:PHYSLB:}, the dynamics of the
superconducting strings has been considered in the framework of
the Nambu-Goto string theory (see, e.g. \cite{Zweibach:2004:CUP}),
and the so called "vortex" theory \cite{Mart-Shel:1997:PRB}.
Similarly to the current of the string loop, the density of the
superconducting electric current $j_s$ is proportional to gradient
of scalar function $\Phi$, which is identified to the phase of the
wave function of the superconducting Cooper pairs
\cite{Ahm-Kag:2005:IJMPD:,Ahm-Fat:2005:IJMPD:}
\beq
j^{\rm s}_\alpha = \frac{2 \hbar n_s e}{m_s} \left(\partial_\alpha \Phi - \frac{2 e}{\hbar c} A_\alpha \right), \label{jELth}
\eeq
where $n_s$ and $m_s$ represent the concentration and mass of the
Cooper pairs, $e$ is the charge of electron and $\hbar$ is the
Planck constant.

In addition to zero resistivity, the superconductors  are
characterized by the existence of the so called Meissner effect
according to which the magnetic field either is  expelled from a
type-I superconductor or penetrates into a type-II superconductor
as an array of vortices. The phase transition between the
superconducting and normal states can be caused by increase of
either temperature or magnetic field. The highest strength of the
magnetic field in a state with given temperature under which a
material remains superconducting is called critical
superconductivity strength. Further, the supercoducting states are
possible only if temperature is lower than the critical
temperature $T_c$, i.e., the temperature of the phase transition
to the supercondicting state. It has to be underlined that the
value of the critical temperature $T_c$ strongly depends on the
pressure, and, e.g. in the neutron star crust, reaches the values
up to $10^9 \sim 10^{10} K$ \cite{Cha-Hae:2008}. The critical
magnetic field at any temperature below the critical temperature
is given by the relation
\beq
B_c \approx B_c(0) \left[1-\left(\frac{T}{T_c}\right)^2\right],
\eeq
where $B_c(0)$ is the critical magnetic field at zero temperature.

Suppose superconducting material is in the external magnetic field
$B<B_c$ at $T > T_c$ and start lowering the temperature. When $T <
T_c$, the medium separates into two phases: superconducting
regions without magnetic flux (flux expulsion) and normal regions
with concentrated strong magnetic field that suppresses
superconductivity. Consequently, magnetic flux can pass through
the superconducting  material, separated into superconducting
regions without magnetic flux and normally conducting regions into
which the magnetic flux is concentrated, and as we discussed the
nature of the nonsuperconducting regions depends on the type of
superconductor as type-I or type-II.

Large-scale, ordered magnetic fields in central parts of accretion
disks around black holes are desirable for explaining e.g.
collimated jet production. However inward advection of vertical
flux in a turbulent accretion disk is problematic if there is an
effective turbulent diffusivity. A new mechanism to resolve this
problem was predicted in \cite{Spr-Uzd:2005:APHJ:}. Turbulent flux
expulsion leads to concentration of the large-scale vertical flux
into small patches of strong magnetic field and because of their
large field strengths, the patches experience higher
angular-momentum loss rate via magnetic braking and winds. As a
result, patches rapidly drift inward, carrying the vertical flux
with them. The accumulated vertical flux aggregates in the central
flux bundle in the inner part of the accretion disk and accretion
flow through the bundle changes its character. It is necessary to
underline that the new phenomenon called \textit{turbulent
diamagnetism} when magnetic field is expelled from regions of
strong turbulence and is concentrated between turbulent cells was
first predicted by Zeldovich \cite{Zel:1956:SOVP:} and Parker
\cite{Par:1963:APJ:}.

Hence, the possibility of appearance of superconductors near the
horizon of black holes, discussed in \cite{Spr-Uzd:2005:APHJ:},
has an important analogy with the turbulence in the accretion
disc, namely, the mechanism of efficient transport of the
large-scale external magnetic flux inward through a turbulent flow
can play a role of superconductivity in the accretion disc and
according to \cite{Spr-Uzd:2005:APHJ:}, suppression of turbulence
by a strong magnetic field is analogous to lifting of
superconductor by an applied magnetic field.

For an electric current-carrying string loop immersed in a
magnetic field there exists a similar effect of the vanishing of
the electric current when the magnetic field is reaching a
critical value. In the theory of the string loops no thermodynamic
features are contained, and the critical phenomena are simply
relating the magnetic field to the electric current (and angular
momentum) parameters of the string loop. To find the critical
value of the magnetic field, let us consider a stationary string
loop located at a fixed radius $r_0$ in the flat spacetime. The
energy per length of the string loop in an asymptotically uniform
magnetic field $B$, given by Eq. (\ref{StringEnergy}) where the
lapse function is reduced to $f(r)=1$ in the flat spacetime, reads
\beq
\frac{E}{r_0} = \mu + \frac{1}{2} \frac{q^2}{r_0^2} + \frac{1}{2} \frac{j^2}{r_0^2} = \mu +\frac{J^2}{r_0^2} + \frac{\Omega J B}{\sqrt{2}} + \frac{B^2 r_0^2}{8}. \label{ensucon}
\eeq
The final expression is the sum of the terms responsible for the string tension, the electric current, the Lorenz force and the energy of the magnetic field, respectively.

If one increases the strength of the background magnetic field $B$ while keeping the string loop at the initial position $r_0$ with fixed energy $E$, the electric current of the string loop has to be modified accordingly. Using the relation (\ref{eletricQJ}), we can write the electric current density $j$ in form corresponding to those related to the superconducting current (\ref{jELth}), and relate the current to the string loop parameters by
\beq
 j = \varphi_{\der \sigma} + A_{\phi} = \pm \sqrt{ - q^2 + 2 E r_0  - 2 \mu r_0^2}.
\eeq
The dependence of the current $j_{\sigma} = \varphi_{\der \sigma}$ (being constant of the motion) on the strength of the magnetic field $B$ is then determined by the relation
\beq
j_{\sigma} = j -\frac{B}{2} r_0^2 . \label{cursucon}
\eeq

According to Eq.~(\ref{cursucon}), the magnitude of the current $j_\sigma$ has to be evidently decreasing with increasing
magnetic field $B$. Therefore, we have to consider the possibility
of disappearance of the electric  $j_\sigma$ current when
magnitude of the magnetic field reaches the critical value. The
critical magnitude of the magnetic field for a string loop located
at a given radius $r_0$ with energy $E$ and charge density $q$
takes the form
\beq
B_{\rm cr} = \pm \frac{2}{r_0^2} \sqrt{ - q^2 + 2 E r_0  - 2 \mu r_0^2}.
\eeq
The presence of the critical magnetic field is provided by the existence of the last term of the Eq.(\ref{ensucon}) representing the energy density of the magnetic field. If the last term in Eq.(\ref{ensucon}) vanishes, the current of the string loop could decrease to arbitrarily small values, but is cannot reach zero value. The presence of the last term of Eq.(\ref{ensucon}) also plays an important role for the benefit of the superconduction -- string loop analogy. The so called Meissner effect of the exclusion of the magnetic field from the superconductor means that the supercurrent generates a magnetic field having strength that is exactly the same as the strength of the external magnetic field. In other words, the term which is proportional to the square of $B$ in Eq.(\ref{ensucon}) can be in the case of the superconducting string loop interpreted as the self-magnetic field generated by the string loop.



\section{String loop acceleration}


Explanation of relativistic jets in Active Galactic Nuclei (AGN)
and microquasars could be one of the most important astrophysical
applications of the string loop model. It is possible because of
the acceleration and fast ejection of the string loop from the
black hole neighbourhood by the transmutation effect, i.e.,
transmission of the energy of the string loop oscillatory motion
in the $x$-direction to the energy of the linear translation
motion along the $y$-direction related to the string loop symmetry
axis
\cite{Lar:1994:CLAQG:,Jac-Sot:2009:PHYSR4:,Stu-Kol:2012:PHYSR4:}.

In the analysis of the acceleration process, it is convenient to use dimensionless coordinates and string loop parameters. We thus make rescaling of the coordinates, $x \rightarrow x/M, y \rightarrow y/M$, and the string loop and background parameters,
\bea
J \rightarrow J / \sqrt{\mu} M, \quad E \rightarrow E / \sqrt{\mu} M, \quad
B \rightarrow BM / \sqrt{\mu}. \label{dmnsnless}
\eea
We will return to the Gaussian units in the Appendix~B.

\begin{figure*}
\includegraphics[width=\hsize]{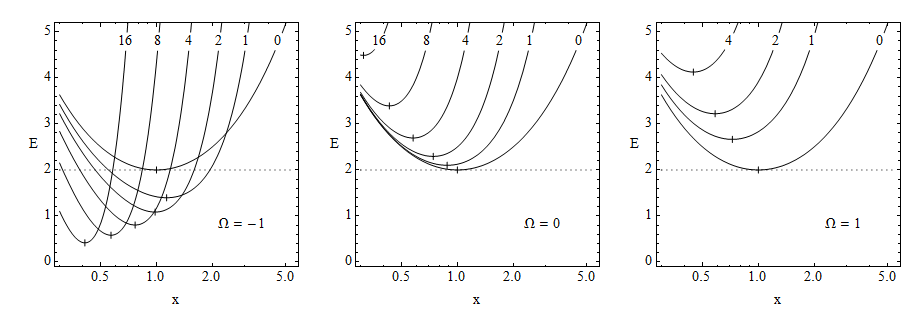}
\caption{String loop effective potential $V_{\rm flat}(x)$ for the flat spacetime with and without an uniform magnetic field. The effective potential is presented for three representative values of the string loop parameter $\Omega \in \{-1,0,1 \}$, keeping constant the current parameter $J=1$, and for various values of the strength of the magnetic field $B$ (denoted by numbers in the plot). Minimum of all the effective potentials is denoted by the small vertical line, the dotted line corresponds to the string loop energy in the non-magnetic case $B=0$.
\label{Emin}}
\end{figure*}

\subsection{Maximal acceleration}

Since the \Schw{} spacetime is asymptotically flat, we have to study the string loop motion in the flat spacetime in order to understand the transmutation process. This enables clear definition of the string loop acceleration process and establishing of maximal acceleration that is available in a given uniform magnetic field. The energy of the string loop (\ref{StringEnergy}) in the flat spacetime with uniform magnetic field $B$, expressed in the Cartesian coordinates, reads
\beq
E^2 = \dot{y}^2 + \dot{x}^2 + V_{\rm flat}(x;B,J,\Omega)  =  E^2_{\mathrm y} + E^2_{\mathrm x}, \label{E2flat}
\eeq
where dot denotes derivative with respect to the affine parameter $\af$ and $V_{\rm flat}(x;B,J,\Omega)$ is the effective potential of the string loop motion in the flat spacetime
\beq
 V_{\rm flat} = \left\{ \frac{B^2 x^3}{8} + \frac{\Omega J B x}{\sqrt{2}} + \left(1 +\frac{J^2}{x^2}\right)x \right\}^2. \label{e-flat}
\eeq

The energy related to the motion in the $x$- and the $y$-directions is given by the relations \cite{Stu-Kol:2012:PHYSR4:}
\beq
  E^2_{\mathrm y} = \dot{y}^2, \quad E^2_{\mathrm x} = \dot{x}^2 + V_{\rm flat}(x;B,J,\Omega) = E^2_{0}.  \label{restenergy}
\eeq
The energy in the $x$-direction $E_{0}$ can be interpreted as an internal energy of the oscillating string, consisting from the potential and kinetic parts; in the limiting case of coinciding minimal and maximal extension of the string loop motion, $x_{\rm i} = x_{\rm o}$, the internal energy has zero kinetic component. The string internal energy can in a well defined way represent the rest energy of the string moving in the $y$-direction in the flat spacetime \cite{Jac-Sot:2009:PHYSR4:,Stu-Kol:2012:PHYSR4:}.

The final Lorentz factor of the transitional motion along the $y$-axis of an accelerated string loop as observed in the asymptotically flat region is determined by the relation \citep{Jac-Sot:2009:PHYSR4:,Stu-Kol:2012:PHYSR4:}
\beq
 \gamma = \frac{E}{E_0} , \label{gamma}
\eeq
where $E$ is the total energy of the string loop having the internal energy $E_{0}$ and moving in the $y$-direction with the velocity corresponding to the Lorentz factor $\gamma$. Clearly, the maximal Lorentz factor of the transitional motion of the string loop is related to the minimal internal energy that can the string loop have, i.e., those with vanishing kinetic energy of the oscillatory motion \cite{Stu-Kol:2012:PHYSR4:}
\beq
 \gamma_{\rm max} = \frac{E}{E_{\rm 0(min)}} \label{gammax}.
\eeq
It should be stressed that rotation of the black hole (or naked
singularity) is not a relevant ingredient of the acceleration of
the string loop motion due to the transmutation effect
\cite{Stu-Kol:2012:PHYSR4:}, contrary to the Blandford---Znajek
effect \cite{Bla-Zna:1977:MNRAS:} usually considered in modelling
acceleration of jet-like motion in AGN and microquasars.

Extremely large values of the gamma factor given by Eq.(\ref{gammax}) can be obtained by setting the initial energy $E$ very large or by adjusting properly the string loop parameters $J,\Omega$ and the magnetic field strength $B$ in order to obtain very low minimal internal energy related to infinity, $E_{\rm 0(min)}$. It is crucial to examine properties of the energy function $E_0(x;J,\Omega,B)$, primarily its minimal allowed value, $E_{\rm 0(min)}(J,\Omega,B)$, given by the local minimum of the effective potential $V_{\rm flat}$ (\ref{e-flat}).

\begin{figure*}
\subfigure[\quad position of the effective potential minima $\tilde{x}_{\rm min}(b)$ ]{\label{Emin1}\includegraphics[width=0.32\hsize]{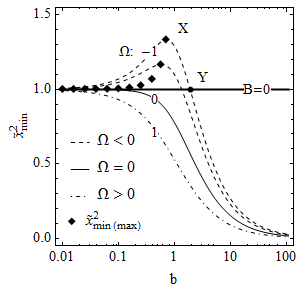}}
\subfigure[\quad value of energy at the minima $\tilde{E}_{\rm (0)min}(b)$ ]{\label{Emin2}\includegraphics[width=0.32\hsize]{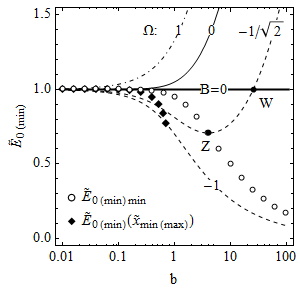}}
\subfigure[\quad properties of the functions $\tilde{x}_{\rm min}(b,\Omega), \tilde{E}_{\rm (0)min}(b,\Omega)$ ]{\label{Emin3}\includegraphics[width=0.32\hsize]{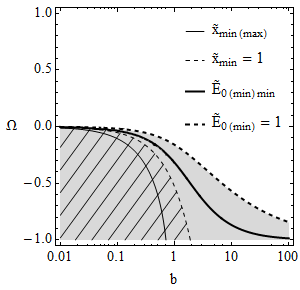}}
\caption{Properties of the effective potential $V_{\rm flat}(x)$ for the string loop dynamics in the flat spacetime with an uniform magnetic field.
Position of the $V_{\rm flat}(x)$ minima $\tx_{\rm (0)min}$ (a), and the minimal energy $\tE_{\rm (0)min}$ (b) are given as function of $b$ for some characteristic values of the string loop parameter $\Omega$. In figure (c), properties of the functions $\tx_{\rm (0)min}(b,\Omega), \tE_{\rm (0)min(b,\Omega)}$ are demonstrated.}
\end{figure*}

Recall that the non-magnetic case, $B=0$, implies quite simple properties of the effective potential $V_{\rm eff}$, as there is only one local minimum with the location and the energy level determined by \cite{Stu-Kol:2012:PHYSR4:}
\beq
 x_{\rm min} = J, \quad E_{\rm 0(min)} = 2 J. \label{nomaglimit}
\eeq
The effective potential $V_{\rm flat}$ is always positive in this case.

Now we have to discuss in detail properties of the effective potential $V_{\rm flat}$ for the string loop dynamics in the homogeneous magnetic field in flat spacetime. It is enough to consider $B>0$ as the situation with opposite sign of $B$ can be described by inversing the sign of $\Omega$. Naturally, also $x>0$. The effective potential diverges for $x \to 0 \,\, (\sim J^2/x)$ and $x \to \infty \,\, (\sim B^2 x^3)$. It is convenient to define a new parameter $b$ reflecting the interaction of the magnetic field and the string loop current, modified space coordinate $\tx$, and modified energy $\tE$, by the relations
\beq
 b = BJ, \quad \tx = \frac{x}{J}, \quad \tE = \frac{E}{2 J}.
\eeq
The zero points of the effective potential are then governed by
\bea
 \tx^{2}_{(z)\pm}(b,\Omega) &=& \frac{ -F \pm \sqrt{F^2 -8b} }{b^2}, \\
  && F = 4 + 2\sqrt{2} b \Omega.
\eea
It is immediately clear that both the solutions $\tx^{2}_{(z)\pm}(b,\Omega)$ can be negative only, being thus physically unrealistic. Therefore, the effective potential is always positive, $V_{\rm flat} > 0$. It has only one local minimum for all values of the string loop parameters, and the magnetic field intensity $B>0$. The extremum, given by $\partial V_{\rm flat} / \partial x = 0$, is located at
\bea
\tx^2_{\rm min} &=& \frac{x^2_{\rm min}}{J^2} = \frac{2\sqrt{2}}{3 b^2}\left\{\sqrt{D}-\left( \sqrt{2} + b \Omega \right) \right\}\\
D &=& 3 b^2 + \left(\sqrt{2} + b \Omega\right)^2
\eea
that exist for all considered values of the parameters $J,B,\Omega$ or $b,\Omega$.
The minimal energy of the string loop in the flat spacetime with the homogeneous magnetic field $B$ is then given by
\beq
\widetilde{E}_{\rm 0(min)} \equiv \frac{E_{\rm 0(min)}}{2 J} = \frac{\sqrt[4]{2}}{3 \sqrt{3}}\,
\frac{\sqrt{D} \left(\sqrt{2} + b \Omega\right) + 6 b^2 - D}{ b \sqrt{\sqrt{D} - \left( \sqrt{2}+ b \Omega \right)}}.
\eeq
Positions of the local minima of the effective potential $\tx^2_{\rm min}$, and the minimal energy $\widetilde{E}_{\rm 0(min)}$, respectively, are illustrated as a function of $b$ for characteristic values of the string loop parameter $\Omega$ in Fig. \ref{Emin1}, and in Fig. \ref{Emin2}, respectively.
The related local extrema of the $\tilde{x}_{\rm min}(b), \tilde{E}_{\rm 0(min)}(b)$ functions are given by the relation
\beq
2 \sqrt{D} \left( b \Omega +\sqrt{2} -\sqrt{D} \right) +b^2 \left(\Omega ^2+3\right)-b \sqrt{D} \Omega +\sqrt{2} b \Omega = 0,
\label{tXextrema}
\eeq
for $\tilde{x}_{\rm min}(b)$, and the relation
\bea
&& b^4 \left(-2 \Omega ^4-3 \Omega ^2+9\right)+2 b^3 \Omega  \left(\sqrt{D} \Omega ^2-\sqrt{2} \left(\Omega ^2-6\right)\right) \nonumber \\
&& +b^2 \left(-9 \sqrt{2} \sqrt{D} +12 \Omega ^2+30\right)+4 b \left(5 \sqrt{2}-3 \sqrt{D} \right) \Omega \nonumber  \\
&& -8 \sqrt{2} \sqrt{D} +16 = 0  \label{tEextrema}
\eea
for $\tilde{E}_{\rm 0(min)}(b)$.

In the limit of $B \to 0$, we find the minimum location at $\tilde{x}_{\rm min} = 1$, and the minimal energy $\tilde{E}_{\rm 0(min)} = 1$, i.e., the values obtained for the empty flat spacetime, given by Eq. (\ref{nomaglimit}). However, these limit values are reached also for a special value of the magnetic field strength $B$, or parameter $b$, in dependence on the other string loop parameter $\Omega$. Therefore, it is relevant to discuss the conditions
\beq
 \tx_{\rm min}(b,\Omega) = 1, \quad \tE_{\rm 0(min)}(b,\Omega) = 1  \label{eq11}
\eeq
in dependence on the interaction of the string loop and the magnetic field expressed by the parameter $b=BJ$. This enables to distinguish qualitatively different string loop configurations from the point of view of the acceleration process. Namely, the condition $\tE_{\rm 0(min)}(b,\Omega) = 1$ separates the string loop configurations enabling efficiency of the acceleration process in the magnetic field to be higher in comparison with those related to the non-magnetic case, from those where the efficiency is lower. The equations (\ref{eq11}) can be expressed in the form
\bea
 -3 b^2 -2\sqrt{2} b \Omega +2\sqrt{2} \sqrt{D}-4 &=& 0, \label{X0good} \\
 2 b^3 \left( \Omega^2 - 1 \right)^2 +2 \sqrt{2} b^2 \Omega \left(3 \Omega^2 + 5\right) &&\nonumber \\
 + 12 b \Omega^2 + b+4 \sqrt{2} \Omega &=& 0.  \label{E0good}
\eea

The numerically determined solutions of the equations governing the local extrema and the "flat limit" values of the position and energy functions, given in terms of the functions $b(\Omega)$, are represented in Fig. \ref{Emin3}. We can see that the solutions exist only in the range of the string loop parameter $\Omega\in\langle-1,0)$.
We denote the solutions of equations (\ref{X0good}) and (\ref{E0good}) by $b_{1\rm(x)}(\Omega)$ and  $b_{1\rm(E)}(\Omega)$, while $b_{E\rm(x)}(\Omega)$ and  $b_{E\rm(E)}(\Omega)$ are referred to solutions of (\ref{tXextrema}) and (\ref{tEextrema}).

For string loop with fixed parameter $\Omega$ the maximal distance
of the $\tilde{x}_{\rm min}(b,\Omega)$ function from the $B=0$
position, $\tilde{x}_{\rm min} = 1$, is given by points in Fig.
\ref{Emin1} and denoted by  $\tilde{x}_{\rm min(max)}$. It is
also useful to give the minimal energy $\tE_{\rm
0(min)}(b,\Omega)$ of the effective potential at the extremal
cases assuming the parameter $\Omega$ fixed and given by the
$b_{E\rm(E)}(\Omega)$ dependence. The results are illustrated by a
sequence of points in Fig. \ref{Emin2} -- the minimal values of
the energy $\tE_{\rm 0(min)}(b,\Omega)$ are denoted by circles,
while the $\tE_{\rm 0(min)}(b,\Omega)$ values obtained at the
maximal distance from  the $\tx_{\rm min(max)}$ point are
denoted by squares.

The loci $\tx_{\rm min}$ of the effective potential $V_{\rm flat}$ local minima depend on the string loop parameter $\Omega$, and the parameter $b$ combining the role of the magnetic field intensity $B$ and the string loop parameter $J$. Keeping the string loop parameters $\Omega$ and $J$ constant, we can obtain a unique position of the minimum $\tx_{\rm min}$ for each magnitude of magnetic field $B>0$ in the case of $\Omega\in\langle0,1\rangle$. Such minima are located at $x_{\rm (min)} < 1$, i.e., closer to the coordinate origin as compared to the non-magnetic ($B=0$) case, see Fig. \ref{Emin1}.

On the other hand, for $\Omega\in\langle-1,0)$ we can obtain a "binary" behavior of the effective potential, if the other parameters are properly tuned. For $b<b_{1\rm(x)}(\Omega)$, we can obtain the same location of the effective potential minimum, $x_{\rm (min)}$, for two different values of parameter $b$. These two different string loop configurations at a given radius will differ in the string loop energy $E$, and have to be located at $\tx_{\rm min} > 1$, i.e., at larger distance from the origin than the minimum of the effective potential in the $B=0$ case. Such "binary" string loop configurations can exist only for subcritical valus of the magnetic parameter
\beq
           b < b_{\rm bin}=4\sqrt{2}/3
\eeq
determined by the condition $b_{1\rm(x)}(-1)=1$ -- see the point Y in Fig. \ref{Emin1}.
The maximal difference of the location of the local minimum of the effective potential from the position $\tx_{\rm min}=1$ corresponding to the non-magnetic case, $B=0$, is given by the point X in Fig. \ref{Emin1}, i.e., by the local maximum of the $\tx_{\rm min}(b)$ function determined by Eq. (\ref{tXextrema}), taken for $\Omega=-1$. For this maximum we obtain
\beq
b_{\rm E(x)}(-1) = 1/\sqrt{2}, \quad \tilde{x}^2_{\rm bin(max)} = \tilde{x}^2_{\rm min}(1/\sqrt{2}) = 4/3.
\eeq


The extremal efficiency of the string loop transmutation process is governed by the minimum of the effective potential at the flat spacetime containing a homogeneous magnetic field, $\tE_{\rm (0)min}(b,\Omega)$, see eq. (\ref{gammax}). If the minimal energy $E_{\rm 0(min)}$ is zero (for example in $J=0, B=0$ case) the maximal possible  acceleration of the string loop diverges, $ \gamma_{\rm max} \rightarrow \infty$.

For positive values of the string loop parameter $\Omega$, range $\Omega\in(0,1\rangle$, the minimal energy function $\tE_{\rm 0(min)}(b)$ is monotonically increasing with increasing parameter $b$ and diverges for $b \to \infty$, see Fig. \ref{Emin2}. Since there is $\tE_{\rm 0(min)} > 1$ for all values of $b>0$, $\Omega\in(0,1\rangle$, such configurations are not advantageous for the string loop acceleration as compared to the non-magnetic case $B=0$.

Configurations $\tE_{\rm 0(min)} < 1$ implying possibility of more efficient string loop acceleration than in the non-magnetic case $B=0$, can exist only for negative values of the parameter $\Omega$. For $\Omega\in(-1,0)$, the minimum energy function $\tE_{\rm 0(min)}(b)$ decreases with increasing $b$ for small enough values of the parameter $b$ reaching a minimum for the magnetic parameter $b_{\rm E(E)}$ (point Z in Fig. \ref{Emin2}) and increases with further increasing of $b$, crossing the $\tE_{\rm 0(min)} = 1$ line at $b_{\rm 1(E)}$ (point W in Fig. \ref{Emin2}) and for larger values of parameter $b$ increases towards infinity. In the special case of $\Omega=-1$, the minimum energy function $\tE_{\rm 0(min)}(b)$ is monotonically decreasing and tends to zero value as $b$ increases to infinity.

For negative values of the string loop parameter,
$\Omega\in\langle-1,0)$, and for a given string loop parameter
$J$, we can always find a properly large values of magnetic field
$B$ to obtain the minimal energy ${E}_{\rm 0(min)}$ of the
effective potential smaller then in the non-magnetic case where
$E_{\rm 0(min)} = 2J$. The ratio of the minimal energy $\tE_{\rm
(0)min}$ considered in the non-magnetic and magnetic cases can be
put arbitrarily close to zero for large enough values of the
parameter $b$. This implies possibility of an extremely efficient
transmutation effect leading to accelerations of the string loop
up velocities corresponding to ultra-high Lorentz factor of the
motion of electric current-carrying string loop with $J>0$ in the
combined Schwarzschild gravitational field and the uniform
magnetic field -- see Fig. \ref{Emin3}.

\begin{figure*}
\includegraphics[width=\hsize]{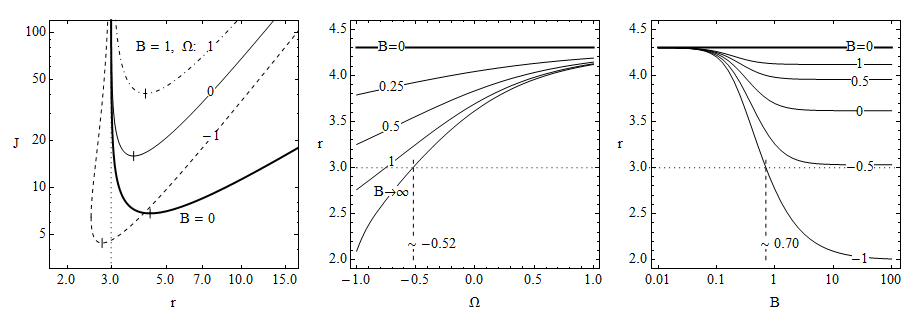}
\caption{ \label{fceJE}
Plots of the $J_{\rm E \mp}(r;B,\Omega)$ function (left figure), position of the extrema of $J_{\rm E \mp}(r;B,\Omega)$ in dependence on the string loop parameter $\Omega$ (central figure), and dependence of the extrema on the strength of the magnetic field $B$ (right figure). \label{picJE}
}
\end{figure*}

\subsection{Acceleration in combined gravitational and magnetic fields}

Clearly, $E_{\rm x}=E_{0}$ and $E_{\rm y}$ are constants of the string loop motion in the flat spacetime and no transmission between these energy modes is possible. However, in vicinity of black holes or naked singularities, the internal kinetic energy of the oscillating string can be transmitted into the kinetic energy of the translational linear motion (or vice versa) due to the chaotic character of the string loop dynamics \cite{Jac-Sot:2009:PHYSR4:,Stu-Kol:2012:PHYSR4:}.

In order to get a strong acceleration, the string loop has to pass the region of strong gravity near the black hole horizon or in vicinity of the naked singularity, where the string loop transmutation effect $E_{\rm x} \leftrightarrow E_{\rm y}$ can occur with maximal efficiency. However, during the acceleration process, all energy of the $E_{\rm x}$ mode cannot be transmitted into the $E_y$ energy mode -- there always remains the inconvertible internal energy of the string, $E_{\rm 0(min)}$, being the minimal energy hidden in the $E_{\rm x}$ energy mode, corresponding to the minimum of the effective potential.

The opposite case corresponds to amplitude amplification of the
oscillations in the $x$-direction and decceleration of the linear
motion in the $y$-direction; in this case the translational
kinetic energy is partially converted to the internal oscillatory
energy of the string. All energy of the transitional ($E_{\rm y}$)
energy mode can be transmitted to the oscillatory ($E_{\rm x}$)
energy mode -- oscillations of the string loop in the
$x$-direction and the internal energy of the string loop will
increase maximally in such a situation, while the string loop will
stop moving in the $y$-direction. We shall focus our attention to
the case of accelerating string loop.

First, we have to discuss the properties of the effective potential of the string loop motion in the combined gravitational and magnetic field. The effective potential is a simple combination of the lapse function $f(r)$ of the spacetime metric, and the effective potential of the string loop motion in the flat spacetime with the uniform magnetic field $V_{\rm flat}$; therefore,
\beq
         V_{\rm eff} = f(r) V_{\rm flat}.
\eeq
The zero point of the effective potential is given by the vanishing of the lapse function $f(r)=0$, i.e., at the Schwarzschild black hole horizon at $r=2$. The divergence occurs at infinity $x \to \infty$, as in the flat spacetime with an homogeneous magnetic field. The local extrema of the effective potential cannot be located off the equatorial plane given by the spherically symmteric spacetime and the uniform magnetic field. In the equatorial plane $y=0$ they are given by the condition \cite{Tur-etal:2013:PHYSR4:}
\bea
 \frac{3}{8} B^2 x^5 - \frac{5}{8} B^2 x^4 + \left(1+\frac{ B J \Omega}{\sqrt{2}}\right) \left( x^3 - x^2\right) - {}
\nonumber\\
 - J^2 x +3 J^2 = 0. \label{extreq}
\eea The local extrema can thus be determined by the condition
related to the angular momentum parameter of the string loop \beq
       J = J_{\rm E \pm}(x;B,\Omega) \equiv \frac{B \Omega x^2 (x-1) \mp \sqrt{G}}{2 \sqrt{2} (x-3)}
\eeq
where
\bea
 G &=& B^2 (x-1)^2 x^2 \Omega^2 + B^2 (x-3) (3 x-5) x^2 \nonumber \\
 && +8 (x-3) (x-1). \label{fceGG}
\eea
There are no zero points of the functions $J_{\rm E \pm}(x;B,\Omega)$ because the condition
\beq
      \frac{1}{8} B^2 x^4 (3x - 5) + x^{2} (x - 1) > 0
\eeq
is satisfied at $x>2$ for all values of the parameter $B$.

The positive branch of the solution $J_{\rm E +}$ is real and positive above the radius of the photon circular geodesic, at $x>3$, for all combinations of parameters $B$ and $\Omega$. For
\beq
\Omega < -\sqrt{\frac{3}{11}} \doteq -0.52,  \quad B > \sqrt{\frac{3 + \sqrt{33}}{18}} \doteq 0.70 \label{bpocon}
\eeq
the solution $J_{\rm E +}$ can be real and positive also below the photon circular geodesic, at $x>2$. Also the solution $J_{\rm E -}$ can be for (\ref{bpocon}) real and positive, but only in the region $2<x<3$.
The behaviour of the functions $J_{\rm E \mp}(x;B,\Omega)$ is demonstrated in Fig. \ref{picJE}.

A given current parameter $J$ represented by a line has possible multiple intersections with the function $J_{\rm E \mp}(x)$ that determine positions of the local extrema of effective potential $V_{\rm eff}(x)$ function in equatorial plane. The local extrema of the $J^2_{\rm E \mp}(x)$ function, given by the condition ${\partial_r J_{\rm E \mp}}~=~0$, enable us to distinguish maxima and minima of effective potential $V_{\rm eff}(x)$. There can exist only local minimum $J_{\rm E \mp (min)}$ for all combination of parameters $B$ and $\Omega$, see Fig. \ref{picJE}.

For the string loop immersed in the combined uniform magnetic
field and the spherically symmetric gravitational field described
by the \Schw{} spacetime, we can have two intersection points of
the $J={\rm const}$ line with the function $J_{\rm E\mp}(x,B,\Omega)$
(maxima and minima of $V_{\rm eff}$) for the parameter $J>J_{\rm E
\mp (min)}$,  one intersection point (inflex point of $V_{\rm
eff}$) for $J=J_{\rm E \mp (min)}$, and none intersection point
for $J<J_{\rm E \mp (min)}$ (no extrema of $V_{\rm eff}$) - the
situation is the same as in the \Schw{} spacetime without magnetic
field \citep{Kol-Stu:2010:PHYSR4:}.

At the minima (maxima) of the effective potential, stable (unstable) equilibrium positions of the string loop occur. Note that energy of the string loop at the stable equilibrium positions governs oscillatory motion around the equilibrium state, but it is not relevant for the maximal acceleration of the string loop in the transmutation process -- the maximal acceleration is given by the local minimum of the effective potential in the flat spacetime \cite{Stu-Kol:2012:JCAP:}.

\begin{figure*}
\includegraphics[width=\hsize]{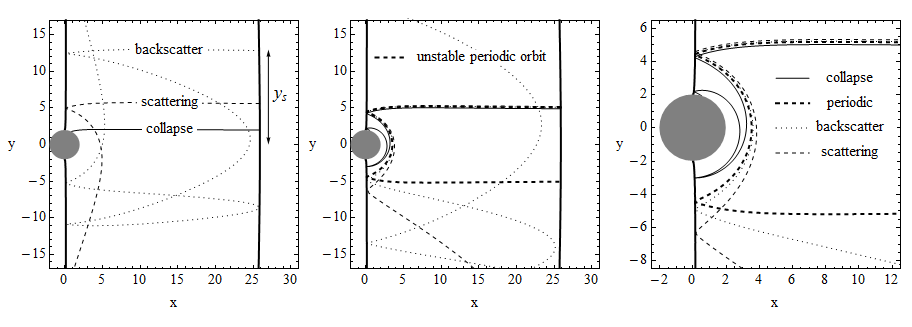}
\caption{Different types of the string loop motion in the combined gravitational Schwarzschild field and uniform magnetic field $B$. The string loop trajectories are represented for appropriately chosen characteristic cases (scattering, backscattering, collapse). The thick line represents boundary of the motion given by the energy boundary function $E=E_{\rm b}(x,y)$, gray is the dynamical region below the black hole horizon.
Sensitivity of the string loop motion to the initial conditions is observed in neighbourhood of the unstable periodic orbit; an example is shown in the second and third figures (being enlargemets of the first one) by the thick dashed line constructed for $y_{\rm s} \doteq 5.059$.
We continuously vary the impact parameter $y_{\rm s}$ and determine by numerical calculations the resulting gamma factor $\gamma(y_{\rm s})$; for the characteristic values of $B$ and $\Omega$ the results are given in Fig. \ref{acceleration1} and Tabs. \ref{tab1}-\ref{tab3}.
\label{schemaACC}}
\end{figure*}

\begin{figure*}
\includegraphics[width=\hsize]{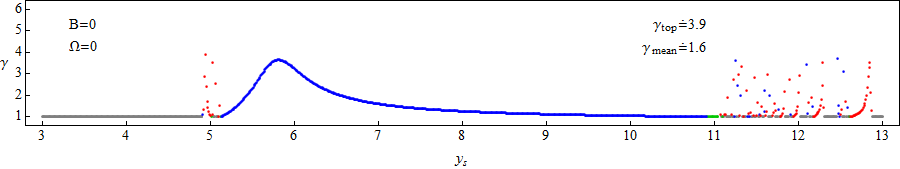}
\includegraphics[width=\hsize]{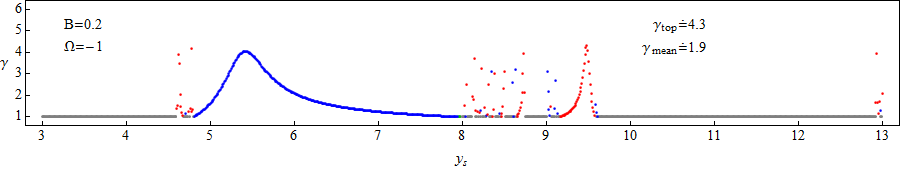}
\includegraphics[width=\hsize]{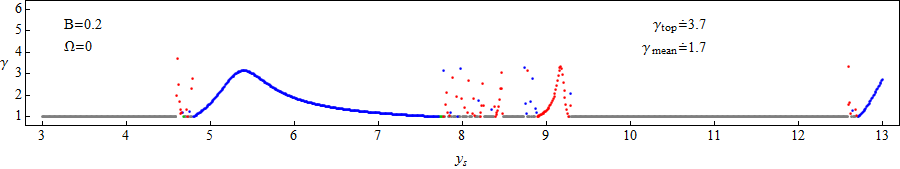}
\includegraphics[width=\hsize]{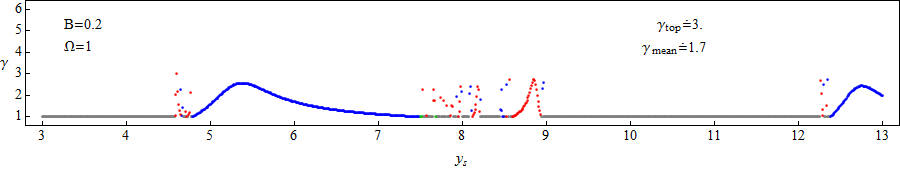}
\caption{\label{acceleration1}
The asymptotic Lorentz factor $\gamma$ obtained due to the transmutation of the string loop energy in the Schwarzschild backgrounds with or without uniform magnetic field $B$ is given for three characteristic values of string loop parameter $\Omega$.
The Lorentz factor $\gamma$ (vertical axis) is calculated for
string loop with energy $E=25$ and current $\JJ=2$, starting from
the rest with varying initial position in the coordinate $\yy_0
\in (0,13) $ (horizontal axis), while the coordinate $\xx_0$ of
the initial position is calculated from Eq. (\ref{StringEnergy}).
Eq. (\ref{gammax}) for the maximal acceleration implies the
limiting gamma factor $\gamma_{\rm max} = 6.25$ for the
non-magnetic case $B=0$. We give the topical Lorentz factor,
$\gamma_{\rm top}$, that is numerically found in the sample, and
also the efficiency of the transmutation effect expressed by the
mean value of the Lorentz factor, $\gamma_{\rm mean}$, in the
sample. The blue (red) colour depicts results of the scattering
(rescattering) of the string loop, the gray colour depicts its
collapse to the black hole. }
\end{figure*}

\begin{figure*}
\includegraphics[width=\hsize]{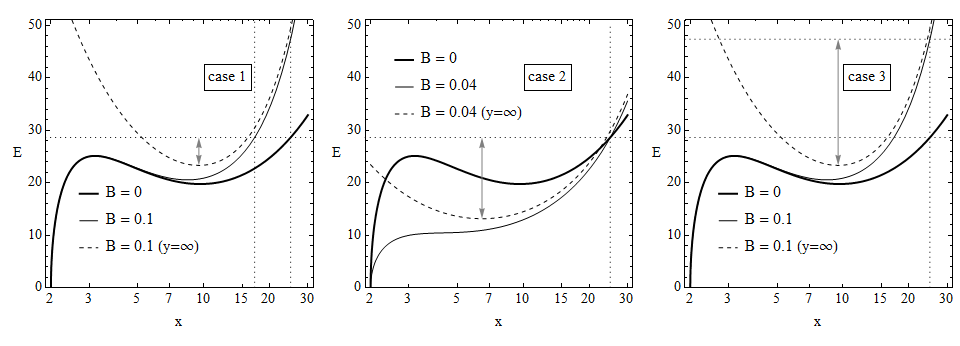}
\caption{The string loop effective potential $V_{eff}(x)$ plotted for various combinations of the parameters $J,B,\Omega$ are illustrated for the flat or \Schw{} spacetime in a way giving the energy that can be used for the string loop acceleration during the transmutation process in dependence on the magnetic field strength $B$.
In the \Schw{} spacetime we use the thick curve for $B=0$, amd the thin curve for $B>0$; the case of the magnetic field in the flat spacetime is dashed.
We will demonstrate the influence of the magnetic field on the string loop acceleration in the three scenarios of modifying the initial conditions because of the increase of $B$: 1) we keep the position $x_i$ but change energy $E$; 2) we keep the energy $E$, but change position $x_i$; 3) we keep the position $x_i$ and the energy $E$, but change the current $J$.
In all figures the string loop parameter $\Omega=0$.
\label{situation}}
\end{figure*}

\subsection{Numerical simulations of the transmutation process}

In the previous section, the maximal possible acceleration of the
string loop has been determined, in dependence on the parameters
$J,B,\Omega$, by finding the minima of the effective potential of
the string loop dynamics in the uniform magnetic field in the flat
spacetime that reflects the asymptotic properties of the combined
Schwarzschild gravitational field and uniform magnetic field.
However, in realistic transmutation processes in the vicinity of
the black hole horizon, the efficiency is usually lower than the
maximally allowed efficiency corresponding to the maximally
accelerated string loop when their oscillatory motion is fully
suppressed.

Due to the chaotic character of the string loop equations of motion, even tiny change in the initial conditions of the motion can change completely character of the string loop trajectory. In order to demonstrate the effect of the magnetic field $B$ on the string loop acceleration, it is useful to compare the set of trajectories with or without magnetic field $B$ in dependence on the string loop parameter $\Omega$ introducing the qualitative differences of the character of the transmutation process, namely in the potential efficiency of the transmutation process reflected by the maximal acceleration determined by $\gamma_{\rm max}$.

The effect of the uniform magnetic field on the string loop
acceleration will be first illustrated by sending the string loop
with fixed current parameters $J$ and $\Omega$ towards the black
hole from rest in the initial position with coordinate $x_{\rm s}$
adjusted in order to have fixed string loop energy $E=25$, but
with freely varying impact parameter $y_{\rm s} \in (0,13)$ --
giving displacement from the equatorial plane, see Fig.
\ref{schemaACC}. Trajectories with large ejection velocity are
assumed to appear for large maximal Lorentz factors,
$\gamma_{\max}$, -- due to Eq. (\ref{gammax}) this should occur
for small values of the string loop current parameter $J$, and
large values of the string loop energy $E$; we will use $J \sim 2$
and $E \sim 25$. Of course, we can test different initials
parameters, but our choice is quite reasonable and illustrative
\cite{Stu-Kol:2012:PHYSR4:}. The current $J$ should be minimized,
but since the minimum of the effective potential $V_{\rm eff}$ is
located at $x_{\rm min} \sim J$, it is difficult for the
trajectories of the string loop with $J<2$ to jump over the black
hole horizon (see Fig. \ref{schemaACC}) and many of the string
loop trajectories will end inside the black hole. The initial
starting point at $x_{\rm s}$ corresponds to an initial stretching
of the string loop -- for larger $x_{\rm s}$, we will start with
larger energy $E$ that also could provide for larger Lorentz
factor $\gamma$ of the string loop far away from the black hole.

For simplicity, we first study the role of the string loop
parameter $\Omega$ on the transmutation process in a fixed
magnetic field with intensity $B=0.2$, considering the
characteristic values of $\Omega = -1, 0, 1$ and compare the
results to the case of vanishing magnetic field $B=0$. The
scattering function $\gamma(y_{\rm s})$, plotted in Fig.
\ref{acceleration1}, demonstrates some regular scattering regions
(for example $y_{\rm s}\in (5.2,11)$ in the $B=0$ case), where the
$\gamma$ depends on $y_{\rm s}$ in a quite regular way, combined with
chaotic scattering regions (chaotic bands \cite{Ott:1993:book:}),
where $\gamma$ depends on $y_{\rm s}$ in a completely chaotic way when
there is no regular prediction of final output from neighbouring
initial positions (for example $y_{\rm s}\in (4.9,5.1)$ in the
$B=0$ case). The reason why there exist such unregular outcomes
from some regions of the initial positions is the presence of
unstable periodic orbits. An illustrative example on the unstable
periodic orbits and behaviour of the string loop trajectories in
its vicinity is presented in Fig. \ref{schemaACC} for our set of
the initial conditions. The results of the study assuming fixed
energy $E$ are reflected in Fig. \ref{acceleration1} by the values
of the Lorentz factors $\gamma_{\rm top}$ and $\gamma_{\rm mean}$, giving
the maximal and mean values of the Lorentz factor obtained from
the considered sample of the initial conditions of the string loop
motion in the given magnetic field. The results confirm
expectation based on the properties of the function
$\gamma_{\rm max}(B,J,\Omega)$ given by Eq. \ref{gammax} that the
transmutation process is most efficient for negative values of the
string loop parameter $\Omega$. Notice that for string loop in the
magnetic field, the factor $\gamma_{\rm top}$ is higher than in the
non-magnetic case $B=0$ only for $\Omega=-1$, but slightly lower
even for $\Omega=0$, and substantially lower for $\Omega=1$. On
the other hand the value of $\gamma_{\rm mean}$ is higher in the
magnetic field for all the three cases of $\Omega$.

Our study seems to be very similar to the problem of chaotic
scattering \cite{Ott:1993:book:} -- the initial vertical
displacement, $y_{\rm s}$, plays the role of the impact parameter,
while we use the resulting string loop Lorentz factor $\gamma$
instead of the scattering angle. One can assume that due to the
chaotic nature of the string loop motion, the numerical
simulations will be able to find the trajectories with Lorentz
factor having almost the maximal value, i.e., $\gamma_{\rm top}
\sim \gamma_{\rm max}$. However, our numerical simulations show
that such an assumption is not generally true, as we obtained at
least $\gamma_{\rm top} \sim 0.5 \gamma_{\rm max}$, or slightly
larger values -- see Fig. \ref{acceleration1} or Tabs.
\ref{tab1}-\ref{tab3}. This is a general effect of the string loop
transmutation process in the field of black holes, related to the
existence of the event horizon capturing the string loop entering
the region of the most efficient transmutation -- in the naked
singularity spacetimes, where the efficient transmutation occurs
in regions containing no event horizon and no capturing of the
string loop occurs, we observe $\gamma_{\rm top} \sim \gamma_{\rm
max}$ frequently \cite{Stu-Kol:2012:JCAP:,Kol-Stu:2013:PHYSR4:}.

Now we have to discuss the influence of the magnitude of the
magnetic field on the transmutation process and compare the
results to those related to the case of $B=0$. The physical reason
for distinction of the transmutations process efficiency in the
\Schw{} black hole without or with magnetic field $B$ comes from
the behaviour of the minimum of the effective potential in the
flat spacetime $E_{\rm 0(min)}$, giving the behaviour of the
accelerated string loop far away from the black hole. For $B=0$ we
have the simple relation $E_{\rm 0(min)}=2J$, while for $B\neq 0$
the value of $E_{\rm 0(min)}$ is modified by the magnetic field
intensity and the string loop parameter $\Omega$. However, the
realistic transmutation process is also strongly influenced by the
presence of the black hole horizon, since the region of minimum of
the effective potential enabling high efficiency of the
transmutation process is close to the black hole horizon
\cite{Stu-Kol:2012:PHYSR4:,Stu-Kol:2012:JCAP:}. Large values of
the Lorentz factor $\gamma$ can be achieved by enlarging the
(initial, and conserved) energy $E$, or by lowering the minimal
string loop energy at infinity, $E_{\rm 0(min)}$, by lowering $J$ --
see Eq. (\ref{gammax}). We can achieve low values of the energy
$E_{\rm 0(min)}$ (see Fig. \ref{Emin}), and hence large acceleration,
by increasing magnitude of the magnetic field $B$, or by using
very low values of the current magnitude $J$ of the string loop
with parameter $\Omega < 0$. However, such string loops minima are
very close to the black hole horizon suppressing thus the
probability of observable acceleration process.

In testing the role of the magnetic field, we have to reflect the problem of the initial conditions that have to be adjusted in order to enable the comparison with the case of $B=0$. The initial conditions for the string loop motion are given by the initial position and the initial speed $x_{\rm s}, y_{\rm s}, \dot{x}_{\rm s}, \dot{y}_{\rm s}$, the internal string loop parameters $E, J, \Omega$, and the external parameter -- intensity of the magnetic field $B$. We cannot choose arbitrarily all the parameters determining the initial conditions, for the string loop starting from the rest the parameters are related by the equation (\ref{StringEnergy}).

 If we want to demonstrate the influence of the magnetic field
on the string loop acceleration process by varying the external
parameter $B$, we have to modify some of the internal string loop
parameters $E,J,\Omega$, or the initial position and the initial
speed. We will discuss three scenarios for the string loop staring
from rest state, distinguished according to what parameter is
varied due to increase of the external parameter $B$, assuming in
all the scenarios the parameter $\Omega$ fixed:
\begin{itemize}
\item[1)] Initial position $x_{\rm s}$ is varied, $E$ and $J$ are fixed
\item[2)] Current parameter $J$ is varied, $x_{\rm s}$ and $E$ are fixed
\item[3)] String energy $E$ is varied, $x_{\rm s}$ and $J$ are fixed.
\end{itemize}
The behaviour of the effective potential in these three scenarios is represented in Fig. \ref{situation}.

\begin{table}[!h]
\begin{center}
\begin{tabular}{l @{\quad} l @{\quad} | @{\quad} c @{\quad} c @{\quad} | @{\quad} c @{\quad} c}
\hline
 & & $x_{\rm 0}$ & $\gamma_{\rm max}$ & $\gamma_{\rm top}$ & $\gamma_{\rm mean}$ \\
\hline \hline
$B=0$ & & $25.8$ & $6.3$ & $3.9$ & $1.6$ \\
\hline
$B=0.1$ & $\Omega=-1$ & $19.5$ & $6.7$ & $4.5$ & $ 1.7$ \\
                & $\Omega=0$ & $18.4$ & $6.2$ & $3.8$ & $ 1.7$ \\
                & $\Omega=1$ & $17.3$ & $5.8$ & $3.3$ & $ 1.6$ \\
\hline
$B=0.2$ & $\Omega=-1$ & $14.7$ & $7.2$ & $4.3$ & $ 1.9$ \\
                & $\Omega=0$ & $13.7$ & $6.2$ & $3.7$ & $ 1.7$ \\
                & $\Omega=1$ & $12.7$ & $5.5$ & $3.0$ & $ 1.7$ \\
\hline
\end{tabular}
\caption{
The characteristic values of the string loop asymptotic Lorentz factor, $\gamma_{\rm top}$ and $\gamma_{\rm mean}$, numerically obtained for the set of trajectories in the acceleration scenario 1 are compared to the maximal Lorentz factor $\gamma_{\rm max}$. In the scenario 1 we keep energy $E=25$ and current $J=2$, while coordinate $x_{\rm s}$ is varied according to increase of the strength of the magnetic field $B$.
} \label{tab1}
\end{center}
\end{table}
\begin{table}[!h]
\begin{center}
\begin{tabular}{l @{\quad} l @{\quad} | @{\quad} c @{\quad} c @{\quad} | @{\quad} c @{\quad} c}
\hline
 & & $J$ & $\gamma_{\rm max}$ & $\gamma_{\rm top}$ & $\gamma_{\rm mean}$ \\
\hline \hline
$B=0$ & & $5.0$ & $2.5$ & $2.5$ & $1.4$ \\
\hline
$B=0.01$ & $\Omega=-1$ & $7.2$ & $1.8$ & $1.8$ & $ 1.3$ \\
                 & $\Omega=0$ & $4.5$ & $2.8$ & $2.7$ & $ 1.4$ \\
                 & $\Omega=1$ & $2.8$ & $4.5$ & $3.2$ & $ 1.5$ \\
\hline
$B=0.02$ & $\Omega=-1$ & $9.4$ & $1.4$ & $1.4$ & $ 1.1$ \\
                 & $\Omega=0$ & $2.3$ & $5.5$ & $4.0$ & $ 1.6$ \\
                 & $\Omega=1$ & $0.6$ & $23.4$ & $4.0$ & $ 1.6$ \\
\hline
\end{tabular}
\caption{
The characteristic values of the string loop asymptotic Lorentz factor, $\gamma_{\rm top}$ and $\gamma_{\rm mean}$, numerically obtained for the set of trajectories in the acceleration scenario 2 are compared to the maximal Lorentz factor $\gamma_{\rm max}$. In the scenario 2 we keep coordinate $x_{\rm s}=25$ and energy $E=25$, while current $J$ is varied according to increase of the strength of the magnetic field $B$.
} \label{tab2}
\end{center}
\end{table}
\begin{table}[!h]
\begin{center}
\begin{tabular}{ l @{\quad} l @{\quad} | @{\quad} c @{\quad} c @{\quad} | @{\quad} c @{\quad} c }
\hline
 & & $E$ & $\gamma_{\rm max}$ & $\gamma_{\rm top}$ & $\gamma_{\rm mean}$ \\
\hline \hline
$B=0$ & & $24.2$ & $6.0$ & $4.0$ & $1.5$ \\
\hline
$B=0.1$ & $\Omega=-1$ & $39.6$ & $10.6$ & $6.6$ & $ 1.7$ \\
                & $\Omega=0$ & $43.0$ & $10.7$ & $6.9$ & $ 1.7$ \\
                & $\Omega=1$ & $46.4$ & $10.8$ & $6.6$ & $ 1.7$ \\
\hline
$B=0.2$ & $\Omega=-1$ & $92.6$ & $26.8$ & $15.5$ & $ 2.2$ \\
                & $\Omega=0$ & $99.4$ & $24.6$ & $14.8$ & $ 2.3$ \\
                & $\Omega=1$ & $106.2$ & $23.3$ & $12.1$ & $ 2.1$ \\
\hline
\end{tabular}
\caption{
The characteristic values of the string loop asymptotic Lorentz factor, $\gamma_{\rm top}$ and $\gamma_{\rm mean}$, numerically obtained for the set of trajectories in the acceleration scenario 3 are compared to the maximal Lorentz factor $\gamma_{\rm max}$. In the scenario 3 we keep coordinate $x_{\rm s}=25$ and current $J=2$, while energy $E$  is varied according to increase of the strength of the magnetic field $B$.
} \label{tab3}
\end{center}
\end{table}

For the first scenario (initial position $x_{\rm s}$ varied with
$B$), the results of the numerical calculations of the Lorentz
$\gamma$ factor are summarized in Tab. \ref{tab1} for two
characteristic values of $B$. As we can see from Tab. \ref{tab1},
the string loop is forced to start closer to the BH horizon for $B
> 0$, if we wish to keep its current $J$ and energy $E$, the
initial position $x_{\rm s}$ decreases with increasing $B$ and
decreases with increasing $\Omega$. There is a slight increase of
the maximal allowed acceleration, $\gamma_{\rm max}$, with
increasing parameter $B$ for $\Omega<0$, while it decreases for
$\Omega>0$, in accord with the discussion of the properties of the
function $E_{\rm 0(min)}(B,J,\Omega)$ in the previous section. For
the string loop with $\Omega=-1$, the topical value of the Lorentz
factor, $\gamma_{\rm top}$, exceeds the value corresponding to the
case $B=0$, while it is lower for $\Omega = 0, 1$. The value of
the $\gamma_{\rm top}$ decreases with increasing $B$ in all the cases
of $\Omega = -1, 0, 1$.

For the second scenario (string loop current parameter $J$ varied
with $B$), the data of the numerical simulations giving the
Lorentz $\gamma$ factors are given in Tab \ref{tab2} for two
characteristic values of $B$ that are by one order smaller than in
the previous case. We have to use smaller values of the magnetic
field $B$, since a critical value $B_{\rm crit}$ of the magnetic
field intensity exists for which the current $J=0$, see section
\ref{analogy}. In this scenario, the parameter $J$ decreases
significantly with increasing $\Omega$ and increasing $B$. We have
found increase of the maximal allowed acceleration $\gamma_{\rm
max}$ in comparison to the case $B=0$ due to decrease of the
current $J$ for $\Omega = 0, 1$, while for $\Omega = -1$ the
current $J$ increased and $\gamma_{\rm max}$ decreased. The value
of $\gamma_{\rm top}$ decreases with increasing $B$ for string loop
with $\Omega = -1$ being lower that in the $B=0$ case, while it
increases with increasing $B$ for $\Omega = 0, 1$ due to the
strong decrease of $J$.

For the third scenario (string loop energy $E$ varied with $B$), the results of the numerical simulations for the Lorentz $\gamma$ factor are summarized in Tab. \ref{tab3} for the same two characteristic values of $B$ as in the first scenario. In comparison to the case of $B=0$, we need increase of the energy $E$ and we observe large increase of the maximal allowed acceleration $\gamma_{\rm max}$ that increases with increasing $\Omega$ for the smaller value of $B=0.1$, but it decreases with increasing $\Omega$ for $B=0.2$. The Lorentz factor topical value $\gamma_{\rm top}$  also demonstrates a substantial increase in comparison to the case $B=0$, especially for the larger magnetic field $B=0.2$. Large $\gamma$ factors are observed for the $\Omega = -1$ case, while for $\Omega = 0, 1$ the increase is smaller; in both cases it is caused by the large increase of the string energy $E$.

\begin{figure*}
\includegraphics[width=\hsize]{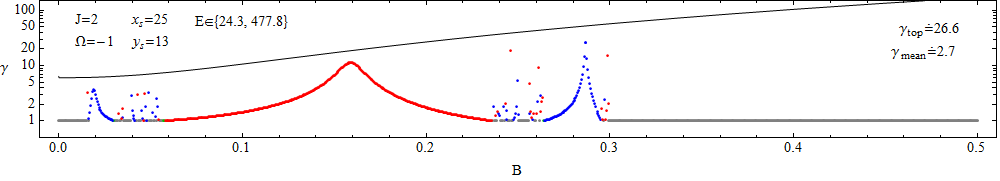}
\includegraphics[width=\hsize]{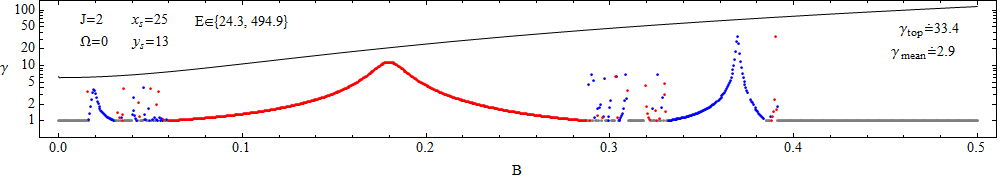}
\includegraphics[width=\hsize]{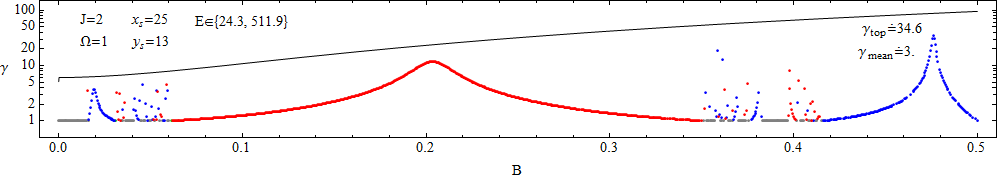}
\caption{\label{aa1} The Lorentz factor $\gamma$ (vertical axis),
calculated for the string loop with current $\JJ=2$, starting from
rest state at a fixed initial position with coordinates $x_0,
y_0$, while the strength of the magnetic field $B$ is varied. The
maximal acceleration given by Eq. (\ref{gammax}) is plotted as the
black curve. The string loop energy $E$ is calculated from Eq.
(\ref{StringEnergy}) and increases with increasing $B$. }
\end{figure*}

In the framework of the third scenario we give also data of the
numerically calculated Lorentz $\gamma$ factors in dependence on
the magnetic field intensity $B$ for the string loop parameters
$\Omega = -1, 0, 1$ and fixed $J=2$, fixed initial position of the
string loop starting at rest, with energy $E$ varied
correspondingly. The data are plotted in Fig. \ref{aa1}; we
observe increasing of the Lorentz $\gamma$ factor with increasing
$B$. We also observe combination of the regions of regular
behaviour of the function $\gamma(B)$, with regions of quite
chaotic character, similar to those occuring in Fig.
\ref{acceleration1}. Notice that $\gamma_{top}$ takes highest
value for the value of $\Omega = 1$ due to highest increase of the
energy $E$ parameter. On the other hand it takes the lowest value
for $\Omega = -1$; moreover, in this case the efficient
acceleration can occur only in the field with $B<0.3$ -- clearly,
for larger values of $B$ the efficient transmutation processes
occur very close to the horizon and the string loop is efficiently
captured by the black hole.

The transmutation between the oscillatory motion and the
transitional accelerated motion can be properly represented by
their distribution in the space of initial states $x_{s}-y_{s}$.
The numerical simulations are giving the resulting Lorentz
$\gamma$ factor for two characteristic values of the magnetic
field $B=0$ and $B=0.2$, two characteristic values of the string
loop parameter $J = 2$ and $J = 11$, and the three characteristic
values of the string loop parameter $\Omega = -1, 0, 1$. The
results are given in Fig. \ref{accelerationXY}, along with the
$E={\rm const}$ levels for the considered cases of the internal
and external parameters of the string loop in the combined
gravitational and magnetic fields. We observe that strong
acceleration and large final Lorentz $\gamma$ factors can be
obtained.

In astrophysically realistic situations, our results are valid up
to the regions distant from the black hole where the magnetic
field can be well approximated as uniform. Clearly, the condition
of magnetic field uniformity assumed to be valid in the black hole
vicinity and reasonably large distance will be violated in regions
very distant from the black hole. Then a crucial question arises
-- how the string loop dynamics will be influenced by decreasing
intensity of the magnetic field in distant regions described by
the flat spacetime. We plan to study this problem in a future
paper. Nevertheless, we can expect that no change will occur in
the flat spacetime far away from the black hole in the energy of
the translational motion $E_{y}$. This energy has to be conserved
since no energy transmutation is possible in the flat spacetime,
and the energy mode $E_y$ is independent on the intensity of the
magnetic field, i.e. the string loop translational velocity along
$y$-axis has to be conserved as well. However, we can expect the
change in the energy of the string loop oscillations in the
$x$-direction since both the total energy, $E$, and the x-mode
energy, $E_{x}$, will be modified by decreasing intensity of the
magnetic field.

Another interesting question for future study arises, if we assume
the uniform magnetic field that is not parallel to the string loop
axis. Then a new force arises that turns the string loop axis to
be parallel with the magnetic field. The non-parallel magnetic
field case also opens up the question of stability of the string
loops in $\Omega>0$ configuration - any perturbation may overturn
the string loop to the $\Omega<0$ configuration that is
energetically more favourable.

\begin{figure*}
\includegraphics[width=\hsize]{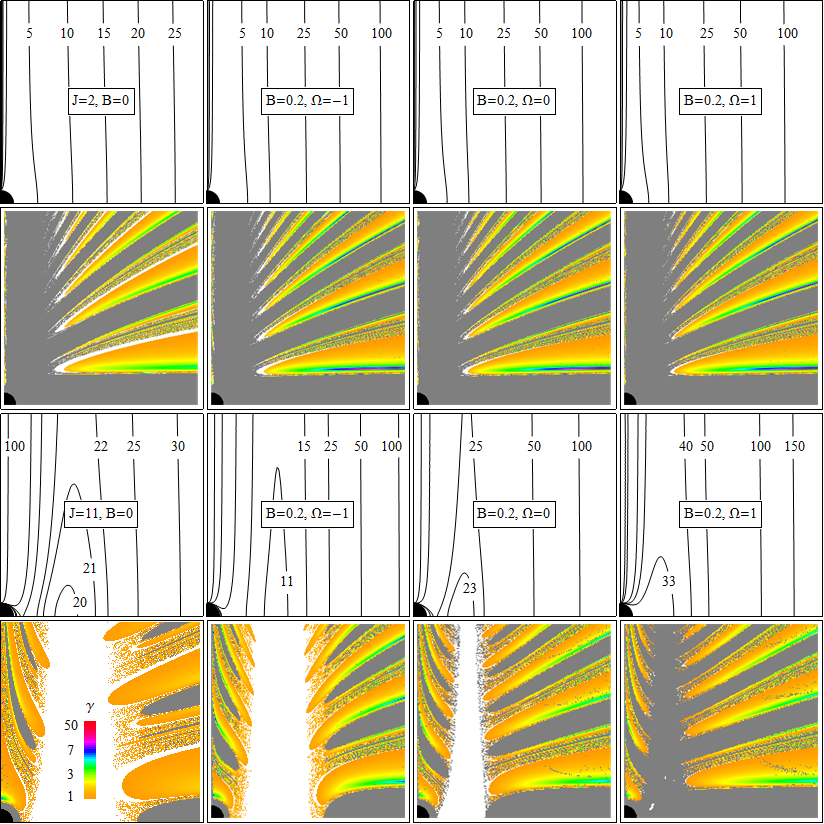}
\caption{ \label{accelerationXY} String loop acceleration in the
\Schw{} spacetime with and without the external homogeneous
magnetic field with $B=0.2$, plotted for various starting points
of the string loop and correspondingly modified energy. The string
loop is starting from the rest at the points from the region
$x_{\rm s} \in (0.1, 30.1), y_{\rm s} \in (0.1, 30.1)$ with the
angular momentum parameter fixed to values of $J=2$ (first two
rows) or $J=11$ (second two rows), but with varying energy $E$
determined by the starting point (the energy levels are
demonstrated in the first, $J=2$, and the third, $J=11$, row).
In the distribution of the Lorentz factor $\gamma$, in the second
row ($J=2$)and the fourth row ($J=11$), we coloured every point
according to the asymptotic Lorentz factor of the translational
string loop motion in the $y$-direction; the code for the colours
is presented in the fourth row. Black colour denotes regions below
the horizon, grey regions correspond to the string loop collapsed
to the black hole. Regions of white colour correspond to string
loop that do not reach "infinity" located at $r=1000$ for given
time $\af=200$ and remain oscillating around the black hole. }
\end{figure*}


\section{Conclusion}

We have investigated acceleration of the electric current-carrying
string loop due to the transmutation process in the gravitational
field of the \Schw{} black hole combined with an
asymptotically uniform magnetic field. We have pointed out a
physical interpretation of the string loop model through the
superconductivity phenomena of plasma in accretions discs. We also
give correspondence of the parameters of the string loop model of
jets to real physical quantities and estimate such quantities in
realistic astrophysical conditions.

In the pure spherically symmetric gravitational field the string
loop dynamics is degenerated, being independent of the string loop
motion constant $\Omega$. In the combined gravitational and
magnetic field that have a common axial symmetry only, the
degeneration is canceled, and the dynamics is strongly dependent
on the parameter $\Omega$. The  effective potential of the string
loop dynamics allows for one stable and one unstable equilibrium
points of the string loop. The maximal acceleration of the string
loop is, however, determined by the minima of the effective
potential of the string loop dynamics in the uniform magnetic
field immersed in the flat spacetime.

The numerical analysis given in Tabs. \ref{tab1}-\ref{tab3} confirms significant acceleration (large $\gamma$ factor) in the cases, where large $\gamma_{\rm max}$ is possible. The string loop acceleration is given by the transmutation process governed by two key ingredients: possibility of the string loop to escape with large ratio of the initial energy $E$ to the minimum energy at infinity $E_{\rm 0(min)}$, and existence of the transmutation region of strong gravitational (magnetic) fields where the chaotic regime of the string loop dynamics occurs and transmission of energy of the oscillatory motion to the energy of the translational motion is possible. It has been proved that presence of the external homogeneous magnetic field $B$ allows the string loop to escape to infinity with large Lorentz $\gamma$ factor, the magnetic field can significantly increase the maximal acceleration given by $\gamma_{\rm max}$.

We have demonstrated that for the positive values of the parameter
$\Omega$, the presence of the magnetic field decreases the
efficiency of the transmutation effect, while it increases the
efficiency for negative values of the parameter $\Omega$. We have
shown that for the intensity of the magnetic field high enough,
the string loop with negative parameter $\Omega$ can be strongly
accelerated up to ultra-relativistic velocities of their
translational motion. Therefore, the string loops accelerated in
the field of magnetized \Schw{} black holes could serve as
an acceptable model of ultra-relativistic jets observed in active
galactic nuclei. The black hole fast rotation is thus not
necessary in the framework of the string loop acceleration model.

One of the most important consequences of our current paper,
considered from the point of view of astrophysics and observable
phenomena, is that the magnetic field substantially increases the
efficiency of the acceleration mechanism of the string loop. The
ultra-relativistic acceleration necessary in modelling the jets
observed in microquasars and active galactic nuclei is shown to be
possible even for non-rotating black holes in the string loop
model. This is a clear opposite to the model of ultra-relativistic
jets based on the Blandford-Znajek process that requires fast
rotating black holes. Therefore, this difference can potentially
give clear signature of relevance of string loop models.

\label{conclusion}

\acknowledgments

The authors would like to express their acknowledgements  for the
Institutional support of the Faculty of Philosophy and Science of
the Silesian University at Opava, the internal student grant of
the Silesian University SGS/23/2013 and the EU grant Synergy
CZ.1.07/2.3.00/20.0071. ZS and MK acknowledge the Albert Einstein
Centre for Gravitation and Astrophysics under the Czech Science
Foundation No. 14-30786G. Warm hospitality that has facilitated
this work to  B.A. by Faculty of Philosophy and Science, Silesian
University in Opava (Czech Republic) and to A.T. and B.A. by the
Goethe University, Frankfurt am Main, Germany is thankfully
acknowledged. The research of B.A. is supported in part by
Projects No. F2-FA-F113, No. EF2- FA-0-12477, and No. F2-FA-F029
of the UzAS and by the ICTP through the OEA-PRJ-29 and the
OEA-NET- 76 projects and by the Volkswagen Stiftung (Grant No. 86
866).

\begin{table*}
\begin{tabular}{c c c c c}
 Quantity & Symbol & Gaussian & Geometrized & Conv. \\
\hline
\\
Length & r & $1$ cm  & $1$ cm  & $1$ \\
$\sigma$ coordinate & $\sigma$ & $1$ cm & $1$ cm & $1$ \\
$\tau$ coordinate & $[\tau]=[c t]$ & $1$ cm & $1$ cm & $1$ \\
Time & t & $1$ s & $2.99\times10^{10}$ cm & $\rm{c}$  \\
Mass & m & $1$ g & $7.42\times10^{-29}$  cm & $\rm{G/c^2}$  \\
Energy & E & $1$ erg & $8.26\times10^{-50}$ cm & $\rm{G/c^4}$ \\
Tension & $\quad \mu \quad$ & $1$ dyn  & $8.26\times10^{-50}$ & $\rm{G/c^4}$ \\
Neutral current & $k\varphi^2_{|a}$ & $ 1 $ $\rm{g\cdot{s^{-1}}}$ & $2.48\times 10^{-39} $ & $\rm{G/c^3}$ \\
Electric current & $j_{\sigma}$ & $ 1 $ statA & $9.59\times 10^{-36} $ & $\rm{\sqrt{G}/c^3}$ \\
Charge & $q$ & $ 1 $ statC & $2.87\times 10^{-25} \rm{cm}$ & $\rm{\sqrt{G}/c^2}$ \\
Charge density & $j_{\tau}$ & $ 1 $ $\rm{statC\cdot{cm^{-1}}}$ & $2.87\times 10^{-25}$ & $\rm{\sqrt{G}/c^2}$ \\
Magnetic field & $B$ & $ 1 $ Gs & $8.16\times 10^{-15} \rm{cm^{-1}}$ & $\rm{\sqrt{G}/c}$ \\
\end{tabular}
\caption{Units and dimensions of the physical quantities of the
string loop in Gaussian (CGS) and geometrized system of units.
\label{tabdim}}
\end{table*}
%

%

\appendix
\section{Estimation of magnetic field intensity in vicinity of black holes}

Throughout the present paper we assume that the external
electromagnetic field
 is weak  in the sense that it is test one
and  does not
effect the background black hole geometry
(\ref{metric}).

The rough estimation
indicates that the local spacetime curvature
produced by the energy of the magnetic
field of intensity $B$ has the order
of $GB^2/c^4$, whereas the
spacetime curvature is of the order of $1/M^2$ near the event
horizon of a black hole with the total mass  $M$. The critical
value $B_M$ of the magnetic field which starts to contribute to
the the spacetime curvature at the reasonable level can be found
from the simple estimation \cite{Fro:2012:PHYSR4:}:
\begin{equation}
 \frac{GB_{M}^2}{c^4}\sim \frac{c^4}{G^2 M^2}\ .
\end{equation}

In other words, the local curvature of the spacetime generated by
the magnetic field $B$ is of the same order or larger than the
gravitational curvature of the black hole spacetime
when~\cite{Fro:2012:PHYSR4:}
\begin{equation}
\label{BBB}
B > B_{M}\approx{c^4\over G^{3/2}
M_{\odot}}\left(\frac{M_{\odot}}{M}\right)\approx
10^{19}{M_{\odot}\over M}~{\textrm{G}}\, .
\end{equation}

Consequently for the test magnetic field $B<<B_M$ when its
influence on the spacetime curvature is totally negligible.
According to the estimations made in~
\cite{Pio-etal:2010:ARXIV:,Fro-Sho:2010:PHYSR4:}, the maximal
strength of the magnetic
 field in the vicinity of the event
horizon of astrophysically
realistic stellar mass black holes or
supermassive black
holes in the Active Galactic nuclei can be
approximated as
\begin{equation}
B \approx 10^8 \textrm{G, \quad for}~M\approx 10 M_{\odot}\ ,
\label{Bstelest}
\end{equation}
\begin{equation}
B \approx 10^4 \textrm{G, \quad for}~M\approx 10^9 M_{\odot}\ .
\label{Bsuperest}
\end{equation}

The main condition $B<<B_M$ is well satisfied for stellar mass and
supermassive black holes both and the magnetic field practically
can not affect
motion of neutral
 particles.

The self magnetic field related to the current-carrying string
loop has been estimated in our preceding paper
\cite{Tur-etal:2013:PHYSR4:}, where we have demonstrated that the
self-magnetic field of the string loop is much smaller than the
external magnetic field, and its influence on the string loop
motion can be abandoned.

\section{Dimensional analysis and estimates of string loop parameters}

In the geometrized units the gravitational constant $G$  and the
speed of light $c$ are taken to be dimensionless units. Their
values in the Gaussian units or so called CGS units are
\bea
 G = 6.67 \times 10^{-8} \,\, \rm{{cm^3}\cdot{g^{-1} \cdot s^{-2}}},  c = 3 \times 10^{10} \,\,
 \rm{{cm}\cdot{s^{-1}}} \ .
\eea
Along with $G$ and $c$, also other world constants are also set to
unity in some unit systems, e.g. Coulomb or electrostatic constant
$k_e = 1/(4\pi\varepsilon_0) = 1$. The conversions of the
fundamental quantities characterizing the string loop motion from
the geometrized units to the Gaussian units are shown in the
Table.~\ref{tabdim}. The table allows to perform transformation
from the geometrized units to the CGS units, and vice versa, for
any dynamical quantity describing the string loop dynamics.

The string loop model enables to apply and compare the derived
solutions for different physical mechanisms to obtain the
estimates of the parameters characterizing the string loop
dynamics. In order to make the estimate of the tension $\mu$
strength, one can use, e.g., the similarity between the role of
the parameter $\mu$ and the Lorentz force acting on a charged
particle in the action governing the string loop dynamics. By
comparison of the forces, one can obtain the tension for the
string loop which is generated by charged particles orbiting in
the vicinity of the black hole. In other words, for such a
comparison the tension of the string loop is considered as an
analogue of the Lorentz force acting on the string loop. Using the
equations (\ref{Sveff}), (\ref{Pveff}) and (\ref{Bstelest}), this
estimate gives in the Gaussian units the tension in order
\beq \mu_{({\rm L})} = 7.2 \times 10^8 {\rm dyn}.
\label{muLorenzEstim} \eeq
For the tension taken as (\ref{muLorenzEstim}), one can find the values of the current and charge densities related to a stable electric current-carrying string loop. According to our previous papers
\cite{Tur-etal:2013:PHYSR4:,Kol-Stu:2010:PHYSR4:,Stu-Kol:2012:JCAP:}, the stable configuration of the string loop implies the values of the dimensionless quantity ${c J^2}/{\mu}$ in the order
\beq \frac{c J^2}{\mu} \sim 10, \label{stableSL}\eeq
i.e., we obtain for the parameter $J^2 \sim 0.24 ~{\rm g \cdot
s^{-1}}$. For the electric current-carrying string loop, i.e. when
the parameters $j_\sigma$ and $j_\tau$ are expressed in units
given by the Table~\ref{tabdim}, one has to set in the definition
of the action (\ref{akceEM}) the parameter $k=1/c^3$ and rewrite
the definition of the current parameter $J$ including the constant
$c$ as
\beq J^2 = \frac{k}{2} \left(j_\sigma^2 + c^2 j_\tau^2\right).
\eeq
In a particular case given by the relation $j_\sigma^2 = c^2
j_\tau^2$, i.e., when $\omega = \pm 1$, one gets the following
estimated values for the charge density and the current of the
string loop
\beq j_{\tau ({\rm L})} \approx 8.5 \times 10^4 {\rm
statC\cdot{cm^{-1}}}, \quad j_{\sigma ({\rm L})} \approx 2.5
\times 10^{15} {\rm ~statA}. \label{estjsjtLorenz}\eeq
Remind, that the estimates given in (\ref{muLorenzEstim}) and
(\ref{estjsjtLorenz}) are obtained on the assumption that the
current carriers by the string loop are the elementary particles
like electrons or protons, i.e. the string loop is generated by
individual charged particles. We call it as a "Lorentz" case, and
it can be considered as the lower limit of our string loop model,
giving minimal estimates of the values of the fundamental
parameters, i.e., the tension, the current, and the charge
density, characterizing the stable, electric current-carrying
string loop.

On the other hand, we can find estimates of the fundamental string loop parameter values related to the so called cosmic strings, giving the upper limit of the application of the string loop model. The cosmic strings are theoretical constructions describing topological defects occurring in the very early universe with ultra-large mass densities due to the spontaneous symmetry breaking of fundamental physical interactions \cite{Vil-She:1994:CSTD:}. The recent summary of the study of the cosmic strings \cite{Bat-Mos:2010:PRD:} demonstrates that the Nambu-Goto type cosmic strings have an upper limit given by the following dimensionless quantity ${G\mu}/{c^4}<2.6\times 10^{-7}$, which implies an upper limit for the tension given by
\beq \mu_{({\rm CS})} < 3.15~\times~10^{42} {\rm dyn}.
\eeq Applying again the relation (\ref{stableSL}) for the tension,
we can find the upper limits for the charge density and the
current carried on the cosmic strings as
\beq j_{\tau ({\rm CS})} < 5.6 \times 10^{21} {\rm
~statC\cdot{cm^{-1}}}, \quad j_{\sigma ({\rm CS})} < 1.7 \times
10^{32} {\rm ~statA}. \label{estjsjtCosmicS} \eeq
Therefore, we can conclude that the minimal range of applicability
of the string loop model for the stable string loop is
approximately determined as
\beq \mu_{({\rm L})} \leq \mu < \mu_{({\rm CS})}, \eeq
\beq j_{\tau ({\rm L})} \leq j_{\tau} < j_{\tau ({\rm CS})},
\qquad j_{\sigma ({\rm L})} < j_{\sigma} \leq j_{\sigma ({\rm
CS})}. \eeq
%


\def\prc{Phys. Rev. C }
\def\pre{Phys. Rev. E }
\def\prd{Phys. Rev. D }
\def\jcap{Journal of Cosmology and Astroparticle Physics }
\def\apss{Astrophysics and Space Science }
\def\mnras{Monthly Notices of the Royal Astronomical Society }
\def\apj{The Astrophysical Journal }
\def\aap{Astronomy and Astrophysics }
\def\actaa{Acta Astronomica }
\def\pasj{Publications of the Astronomical Society of Japan }
\def\apjl{Astrophysical Journal Letters }
\def\pasa{Publications Astronomical Society of Australia }
\def\nat{Nature }
\def\physrep{Physics Reports }

\bibliographystyle{h-physrev}
\bibliography{reference}

\begin{thebibliography}{10}

\bibitem{Jac-Sot:2009:PHYSR4:}
T.~{Jacobson} and T.~P. {Sotiriou},
\newblock \prd {\bf 79}, 065029 (2009), 0812.3996.

\bibitem{Sem-Ber:1990:ASS:}
V.~S. {Semenov} and L.~V. {Bernikov},
\newblock \apss {\bf 184}, 157 (1991).

\bibitem{Chri-Hin:1999:PhRvD:}
M.~{Christensson} and M.~{Hindmarsh},
\newblock \prd {\bf 60}, 063001 (1999), astro-ph/9904358.

\bibitem{Sem-Dya-Pun:2004:Sci:}
V.~{Semenov}, S.~{Dyadechkin}, and B.~{Punsly},
\newblock Science {\bf 305}, 978 (2004), astro-ph/0408371.

\bibitem{Spr:1981:AA:}
H.~C. {Spruit},
\newblock \aap {\bf 102}, 129 (1981).

\bibitem{Cre-Stu:2013:PhRvE:}
C.~{Cremaschini} and Z.~{Stuchl{\'{\i}}k},
\newblock \pre {\bf 87}, 043113 (2013).

\bibitem{Cre-Stu-Tes:2013:PlasmaPhys:}
C.~{Cremaschini}, Z.~{Stuchl{\'{\i}}k}, and M.~{Tessarotto},
\newblock Physics of Plasmas {\bf 20}, 052905 (2013).

\bibitem{Cre-Stu:2014:PlasmaPhys:}
C.~Cremaschini and Z.~Stuchlík,
\newblock Physics of Plasmas (1994-present) {\bf 21},  (2014).

\bibitem{Kol-Stu:2013:PHYSR4:}
M.~{Kolo{\v s}} and Z.~{Stuchl{\'{\i}}k},
\newblock \prd {\bf 88}, 065004 (2013), 1309.7357.

\bibitem{Kol-Stu:2010:PHYSR4:}
M.~{Kolo{\v s}} and Z.~{Stuchl{\'{\i}}k},
\newblock \prd {\bf 82}, 125012 (2010), 1103.4005.

\bibitem{Stu-Kol:2012:JCAP:}
Z.~{Stuchl{\'{\i}}k} and M.~{Kolo{\v s}},
\newblock \jcap {\bf 10}, 8 (2012), 1309.6879.

\bibitem{Lar:1994:CLAQG:}
A.~L. {Larsen},
\newblock Classical and Quantum Gravity {\bf 11}, 1201 (1994), hep-th/9309086.

\bibitem{Fro-Lar:1999:CLAQG:}
A.~V. {Frolov} and A.~L. {Larsen},
\newblock Classical and Quantum Gravity {\bf 16}, 3717 (1999), gr-qc/9908039.

\bibitem{Stu-Kol:2014:PHYSR4:}
Z.~{Stuchl{\'{\i}}k} and M.~{Kolo{\v s}},
\newblock \prd {\bf 89}, 065007 (2014), 1403.2748.

\bibitem{Stu-Kol:2012:PHYSR4:}
Z.~{Stuchl{\'{\i}}k} and M.~{Kolo{\v s}},
\newblock \prd {\bf 85}, 065022 (2012), 1206.5658.

\bibitem{Stu-Sche:2013:CLAQG:}
Z.~{Stuchl{\'{\i}}k} and J.~{Schee},
\newblock Classical and Quantum Gravity {\bf 30}, 075012 (2013).

\bibitem{Bla-Zna:1977:MNRAS:}
R.~D. {Blandford} and R.~L. {Znajek},
\newblock \mnras {\bf 179}, 433 (1977).

\bibitem{Pun:2001:Springer:}
B.~{Punsly},
\newblock {\em {Black hole gravitohydromagnetics}} (, 2001).

\bibitem{Gar-etal:2010:ASTRA:}
J.~{Gariel}, M.~A.~H. {MacCallum}, G.~{Marcilhacy}, and N.~O. {Santos},
\newblock \aap {\bf 515}, A15 (2010).

\bibitem{Gar-Mar-San:2013:ApJ:}
J.~{Gariel}, G.~{Marcilhacy}, and N.~O. {Santos},
\newblock \apj {\bf 774}, 109 (2013), 1303.6474.

\bibitem{Koo-Bic-Kun:1999:PASA:}
M.~{de Kool}, G.~V. {Bicknell}, and Z.~{Kuncic},
\newblock \pasa {\bf 16}, 225 (1999).

\bibitem{Mil-etal:2006:NATUR:}
J.~M. {Miller} {\em et~al.},
\newblock \nat {\bf 441}, 953 (2006), astro-ph/0605390.

\bibitem{Fro:2012:PHYSR4:}
V.~P. {Frolov},
\newblock \prd {\bf 85}, 024020 (2012), 1110.6274.

\bibitem{Prasanna:1980:RDNC:}
A.~R. {Prasanna},
\newblock Nuovo Cimento Rivista Serie {\bf 3}, 1 (1980).

\bibitem{Fro-Sho:2010:PHYSR4:}
V.~P. {Frolov} and A.~A. {Shoom},
\newblock \prd {\bf 82}, 084034 (2010), 1008.2985.

\bibitem{Kov-Stu-Kar:2008:CAQG:}
J.~{Kov{\'a}{\v r}}, Z.~{Stuchl{\'{\i}}k}, and V.~{Karas},
\newblock Classical and Quantum Gravity {\bf 25}, 095011 (2008), 0803.3155.

\bibitem{Kov-etal:2010:CAQG:}
J.~{Kov{\'a}{\v r}}, O.~{Kop{\'a}{\v c}ek}, V.~{Karas}, and
  Z.~{Stuchl{\'{\i}}k},
\newblock Classical and Quantum Gravity {\bf 27}, 135006 (2010), 1005.3270.

\bibitem{Kop-etal:2010:ApJ:}
O.~{Kop{\'a}{\v c}ek}, V.~{Karas}, J.~{Kov{\'a}{\v r}}, and
  Z.~{Stuchl{\'{\i}}k},
\newblock \apj {\bf 722}, 1240 (2010), 1008.4650.

\bibitem{Wagh-Dadhich:1989:PR:}
S.~M. {Wagh} and N.~{Dadhich},
\newblock \physrep {\bf 183}, 137 (1989).

\bibitem{Las-etal:2014:PYSR4:}
J.-P. {Lasota}, E.~{Gourgoulhon}, M.~{Abramowicz}, A.~{Tchekhovskoy}, and
  R.~{Narayan},
\newblock \prd {\bf 89}, 024041 (2014), 1310.7499.

\bibitem{Li-Xin:2000:PHYSR4:}
L.-X. {Li},
\newblock \prd {\bf 61}, 084016 (2000), astro-ph/9902352.

\bibitem{Kov:2013:EPJP:}
J.~{Kov{\'a}{\v r}},
\newblock European Physical Journal Plus {\bf 128}, 142 (2013).

\bibitem{Stu:1983:BULAI:}
Z.~{Stuchlik},
\newblock Bulletin of the Astronomical Institutes of Czechoslovakia {\bf 34},
  129 (1983).

\bibitem{Stu-Hle:1999:PHYSR4:}
Z.~{Stuchl{\'{\i}}k} and S.~{Hled{\'{\i}}k},
\newblock \prd {\bf 60}, 044006 (1999).

\bibitem{Stu-Sla-Kov:2009:CLAQG:}
Z.~{Stuchl{\'{\i}}k}, P.~{Slan{\'y}}, and J.~{Kov{\'a}{\v r}},
\newblock Classical and Quantum Gravity {\bf 26}, 215013 (2009), 0910.3184.

\bibitem{Stu-Sche:2011:JCAP:}
Z.~{Stuchl{\'{\i}}k} and J.~{Schee},
\newblock \jcap {\bf 9}, 18 (2011).

\bibitem{Tur-etal:2013:PHYSR4:}
A.~{Tursunov}, M.~{Kolo{\v s}}, B.~{Ahmedov}, and Z.~{Stuchl{\'{\i}}k},
\newblock \prd {\bf 87}, 125003 (2013).

\bibitem{Lar:1993:CLAQG:}
A.~L. {Larsen},
\newblock Classical and Quantum Gravity {\bf 10}, 1541 (1993), hep-th/9304086.

\bibitem{GOTO:1971:POTP}
T.~{Got{\= o}},
\newblock Progress of Theoretical Physics {\bf 46}, 1560 (1971).

\bibitem{Car-Ste:2004:PHYSR4:}
B.~{Carter} and D.~A. {Steer},
\newblock \prd {\bf 69}, 125002 (2004), hep-th/0307161.

\bibitem{Wald:1974:PHYSR4:}
R.~M. {Wald},
\newblock \prd {\bf 10}, 1680 (1974).

\bibitem{Mis-Tho-Whe:1973:Gra:}
C.~W. {Misner}, K.~S. {Thorne}, and J.~A. {Wheeler},
\newblock {\em {Gravitation}} (, 1973).

\bibitem{Wit:1985:NuclPhysB:}
E.~{Witten},
\newblock Nuclear Physics B {\bf 249}, 557 (1985).

\bibitem{Vil-She:1994:CSTD:}
A.~{Vilenkin} and E.~P.~S. {Shellard},
\newblock {\em {Cosmic Strings and Other Topological Defects}} (, 1995).

\bibitem{Kim:1999:JKPS}
H.-C. {Kim}, Y.~{Kim}, and B.~K. {Lee},
\newblock Journal of Korean Physical Society {\bf 35}, 649 (1999).

\bibitem{Mart-Shel:1997:PRB}
C.~J.~A.~P. {Martins} and E.~P.~S. {Shellard},
\newblock \prb {\bf 56}, 10892 (1997), cond-mat/9607093.

\bibitem{Car:1990:PHYSLB:}
B.~{Carter},
\newblock Physics Letters B {\bf 238}, 166 (1990), hep-th/0703023.

\bibitem{Zweibach:2004:CUP}
B.~{Zwiebach},
\newblock {\em {A First Course in String Theory}} (, 2004).

\bibitem{Ahm-Kag:2005:IJMPD:}
B.~J. {Ahmedov} and V.~G. {Kagramanova},
\newblock International Journal of Modern Physics D {\bf 14}, 837 (2005),
  gr-qc/0608017.

\bibitem{Ahm-Fat:2005:IJMPD:}
B.~J. {Ahmedov} and F.~J. {Fattoyev},
\newblock International Journal of Modern Physics D {\bf 14}, 817 (2005),
  astro-ph/0607671.

\bibitem{Cha-Hae:2008}
N.~Chamel and P.~Haensel,
\newblock Living Reviews in Relativity {\bf 11} (2008).

\bibitem{Spr-Uzd:2005:APHJ:}
H.~C. {Spruit} and D.~A. {Uzdensky},
\newblock \apj {\bf 629}, 960 (2005), astro-ph/0504429.

\bibitem{Zel:1956:SOVP:}
Y.~B. {Zeldovich},
\newblock Sov. Phys. JETP {\bf 51}, 460 (1957).

\bibitem{Par:1963:APJ:}
E.~N. {Parker},
\newblock \apj {\bf 138}, 552 (1963).

\bibitem{Ott:1993:book:}
E.~{Ott},
\newblock {\em {Chaos in dynamical systems}} (, 1993).

\bibitem{Pio-etal:2010:ARXIV:}
M.~Y. {Piotrovich}, N.~A. {Silant'ev}, Y.~N. {Gnedin}, and T.~M.
  {Natsvlishvili},
\newblock ArXiv e-prints  (2010), 1002.4948.

\bibitem{Bat-Mos:2010:PRD:}
R.~{Battye} and A.~{Moss},
\newblock \prd {\bf 82}, 023521 (2010), 1005.0479.

\end{thebibliography}


\end{document}